\UseRawInputEncoding

\documentclass[a4paper,fleqn]{cas-sc}

\usepackage[authoryear]{natbib}
\usepackage{float}
\def\tsc#1{\csdef{#1}{\textsc{\lowercase{#1}}\xspace}}
\tsc{WGM}
\tsc{QE}
\usepackage[utf8]{inputenc}
\usepackage{multirow}

\begin{document}
\let\WriteBookmarks\relax
\shorttitle{Real-Time Cognitive Training Adaptation based on Eye-Tracking and Physiological Data in Virtual Reality}    

\shortauthors{Szczepaniak et al.}  

\title [mode = title]{Your Eyes Controlled the Game: Real-Time Cognitive Training Adaptation based on Eye-Tracking and Physiological Data in Virtual Reality}  


\tnotetext[1]{supported by UKRI Centre for Doctoral Training in Socially Intelligent Artificial Agents, Grant Number EP/SO2266X/1} 


\author[1]{Dominik Szczepaniak}[orcid=0009-0009-6198-0253]

\cormark[1]


\ead{d.szczepaniak.1@research.gla.ac.uk}


\affiliation[1]{organization={Social AI CDT, University of Glasgow},
            %addressline={}, 
            city={Glasgow},
%          citysep={}, % Uncomment if no comma needed between city and postcode
            %postcode={}, 
            %state={},
            country={United Kingdom}}

\author[2]{Monika Harvey}[orcid=0000-0003-1694-1174
]

% Footnote of the second author

% Email id of the second author
\ead{monika.harvey@glasgow.ac.uk}



\affiliation[2]{organization={School of Psychology \& Neuroscience, University of Glasgow},
            %addressline={}, 
            city={Glasgow},
%          citysep={}, % Uncomment if no comma needed between city and postcode
            %postcode={}, 
            %state={},
            country={United Kingdom}}

\author[3]{Fani Deligianni}[orcid=0000-0003-1306-5017
]

% Footnote of the second author
\cormark[1]

% Email id of the second author
\ead{fani.deligianni@glasgow.ac.uk}



\affiliation[3]{organization={School of Computing Science, University of Glasgow},
            %addressline={}, 
            city={Glasgow},
%          citysep={}, % Uncomment if no comma needed between city and postcode
            %postcode={}, 
            %state={},
            country={United Kingdom}}
\cortext[1]{Corresponding authors}


\begin{abstract}
Cognitive training for sustained attention and working memory is vital across domains relying on robust mental capacity such as education or rehabilitation. Adaptive systems are essential, dynamically matching difficulty to user ability to maintain engagement and accelerate learning. Current adaptive systems often rely on simple performance heuristics or predict visual complexity and affect instead of cognitive load. This study presents the first implementation of real-time adaptive cognitive load control in Virtual Reality cognitive training based on eye-tracking and physiological data. We developed a bidirectional LSTM model with a self-attention mechanism, trained on eye-tracking and physiological (PPG, GSR) data from 74 participants. We deployed it in real-time with 54 participants across single-task (sustained attention) and dual-task (sustained attention + mental arithmetic) paradigms. Difficulty was adjusted dynamically based on participant self-assessment or model's real-time cognitive load predictions. Participants showed a tendency to estimate the task as too difficult, even though they were objectively performing at their best. Over the course of a 10-minute session, both adaptation methods converged at equivalent difficulty in single-task scenarios, with no significant differences in subjective workload or game performance. However, in the dual-task conditions, the model successfully pushed users to higher difficulty levels without performance penalties or increased frustration, highlighting a user tendency to underestimate capacity under high cognitive load. Findings indicate that machine learning models may provide more objective cognitive capacity assessments than self-directed approaches, mitigating subjective performance biases and enabling more effective training by pushing users beyond subjective comfort zones toward physiologically-determined optimal challenge levels.

\end{abstract}


% Keywords

\maketitle
\section{Introduction}

\subsection{Cognitive Training: From Traditional Methods to Modern Challenges}

Cognitive training, or the structured practice of specific cognitive domains to maintain or improve mental functioning, has emerged as a critical intervention strategy. While traditionally prioritized to address the growing prevalence of cognitive impairment in aging populations (\cite{brugada2022enhance}), the relevance of these interventions extends to other age groups, with recent meta-analyses confirming the efficacy of cognitive training interventions for improving executive functions in children (\cite{cassidy2024training}). Traditional approaches, predominantly employing paper-and-pencil methods or screen-based computerized cognitive training (CCT), have demonstrated improvements in specific cognitive domains. For instance, CCT programs have shown gains in memory, reasoning, and speed-of-processing that can be maintained for 5 (\cite{willis2006long}) or even 10 years (\cite{rebok2014ten}). Furthermore, recent evidence highlights that multidomain training strategies can yield broader benefits, successfully transferring to untrained attentional and executive functions in healthy older adults (\cite{motes2020multidomain}).

\subsection{The Promise of Virtual Reality for Cognitive Training}

VR environments, however, offer immersive environments (\cite{brugada2022enhance}) and a unique combination of experimental control and user engagement that traditional methods cannot achieve. The sense of presence and immersion in the virtual environment, together with the naturalistic interaction with the environment provide a higher degree of ecological validity than current screen-based or pen-and-paper solutions (\cite{brugada2022enhance}).

In fact, recent systematic reviews and meta-analyses provide compelling evidence for VR's effectiveness in cognitive rehabilitation. For example, (\cite{yu2023effect}) showed that VR training had an overall positive effect on cognitive flexibility, global cognitive function, attention, and short-term memory compared to the control groups , with effect sizes suggesting meaningful clinical improvements. Moreover, VR-based interventions are a potential at-home therapy option, offering individuals with cognitive decline an engaging and accessible means of cognitive improvement (\cite{moulaei2024efficacy}), addressing the critical need for scalable, accessible cognitive interventions.

This at-home flexibility also increases accessibility for patients with mobility issues, or those far from clinics, ensuring that more individuals can benefit from advanced rehabilitation techniques. This is particularly crucial given the growing need for scalable cognitive interventions that can reach diverse populations without requiring specialized clinical facilities.

\subsection{Sustained Attention and Cognitive Load Assessment}

Sustained attention, defined as the ability to maintain focused cognitive activity over extended periods, is a fundamental cognitive capacity underlying successful performance across numerous real-world tasks. Cognitive impairments of processing speed, sustained attention, and working memory are frequently reported and impact negatively on activities of daily living and quality of life (\cite{johansen2024virtual}). The importance of sustained attention extends beyond clinical populations, as it serves as a cornerstone for academic achievement, occupational performance, and daily functioning across the lifespan.

In particular, sustained attention supports ongoing and optimized performance in activities demanding high levels of focus and active cognitive engagement. It involves the capacity to stay vigilant and attentive throughout monotonous or repetitive tasks, necessitating focus on pertinent stimuli while disregarding distractions and evolving circumstances (\cite{cohen2014focused}). Two primary constraints are crucial for creating sustained attention tasks: capacity (the volume of information being attended to) and selectivity (the extent to which unattended information is still processed) (\cite{cochrane2020load}). When the intrinsic (task-specific) load diminishes, the influence of distractors grows because additional processing capacity becomes accessible (\cite{rao2020predicting}). Consequently, the two principal hazards in handling cognitive load within a sustained attention task are that the design might be overly simplistic, permitting distractors to divert attention from the task, or that the intrinsic load could be excessively elevated, surpassing the participant's attentional capacity and thereby impairing performance. A typical experimental setup in the realm of sustained attention entails reacting to chosen stimuli during extended intervals of inactivity, such as observing a sequence of letters and responding solely to a (rare) target (\cite{hamilton2023sustained}). Owing to these prolonged inactive periods, this approach is susceptible to generating boredom. This, in sequence, leads to participants withdrawing from the task, directing greater attention toward distractions, and experiencing diminished task performance accordingly. In contrast, ongoing assessment of cognitive load facilitates real-time adjustments to the assigned task load, ideally resulting in a task that is neither overly demanding nor insufficiently stimulating \cite{liu2017multisubject}.

Mental arithmetic can serve as an ideal secondary task for cognitive load manipulation: it adds a new dimension particularly suitable for creating controlled cognitive load conditions while maintaining ecological validity, as arithmetic operations are commonly encountered in daily life situations (\cite{mcphee2022dual}).

\subsection{Limitations of EEG-Based Approaches for Cognitive Load Assessment}

Due to its high temporal resolution and direct measurement of neural activity, electroencephalography (EEG) has been extensively used for cognitive load assessment. For instance, \cite{tremmel2019estimating} used EEG to predict cognitive load in VR and obtained an accuracy of 81.1\% on a binary classification and 63.9\% on a 33.3\% chance classification. 
However, there are significant limitations that constrain its practical application. Most experiments on passive BCI use a very controlled approach, which naturally limits the range of real-world conditions they reflect. While this control is necessary to ensure the psychophysiological validity of the mental state detection, the results lack ecological validity, and can not be generalized to other contexts (\cite{muhl2014eeg}).

Additionally, EEG systems also present practical challenges for deployment in interactive VR environments: electrode placement and motion artifacts can significantly compromise signal quality. Moreover, incorporating EEG headsets increases setup time and adds to the fatigue experienced by the participants (\cite{hanzal2023investigation}).

\subsection{Eye-tracking and physiological data as suitable cognitive load assessors/predictors.}

Recent research has demonstrated that eye-tracking metrics serve as an effective means for evaluating cognitive load, proving less susceptible to noise compared to EEG (\cite{aygun2022investigating}) and simpler to implement (as many VR headsets include built-in eye-trackers). Indeed, indicators such as pupil diameter, gaze fixation, blink duration, and saccadic movements have been extensively linked to cognitive load (\cite{bitkina2021ability}). In a study by \cite{hebbar2022cognitive}, a blend of EEG and eye-tracking data was employed during a VR flight simulator task to investigate cognitive load indicators. Their findings revealed that changes in pupil diameter, fixation rate, and gaze distribution were associated with cognitive load, as indicated by EEG activity in that context. Likely owing to their limited sample of 12 participants, they did not attempt to predict cognitive load via eye-tracking metrics, so no accuracy is available.

Multiple investigations have sought to estimate cognitive load through Machine Learning (ML) techniques utilizing eye-tracking approaches.  \cite{RN10} presented a dataset involving 57 participants and eye-tracking recordings at a 1000Hz sampling rate. Employing regression analysis solely with eye-tracking data, they achieved R-squared values of 0.362 for XGBoost and 0.352 for Random Forest Regression via a leave-one-out validation approach. A separate study on predicting cognitive load among drivers (\cite{shojaeizadeh2019detecting}) reached an accuracy of 79\% with a Random Forest Classifier on an 80/20 data split across 48 subjects, with pupillary responses emerging as the strongest model contributors. In yet another investigation (\cite{skaramagkas2021cognitive}), eye-tracking-derived metrics were used to perform binary classification of cognitive load based on NASA-TLX scores (a self-report questionnaire for subjective workload assessment (\cite{hart1988development})). The researchers opted against randomizing the data, citing its temporal characteristics as unsuitable for such a method. Instead, they allocated 81\% for training, 9\% for validation, and 10\% for testing, conducting 10-fold cross-validation and attaining a peak accuracy of 88\%.

Certain other studies have concentrated on gauging cognitive workload using physiological indicators like heart rate and galvanic skin response (GSR). For instance, \cite{he2022classification} integrated eye-tracking with heart rate and GSR data to examine cognitive load in drivers. By combining these modalities, the authors achieved a top accuracy of 97.8\% using a Random Forest Classifier, noting that physiological metrics enhanced model accuracy by 34.5\%. Since physiological data can be readily captured via everyday wearable devices such as smartwatches, they hold potential to support the creation of reliable machine learning models for real-time cognitive workload monitoring.

The most robust dataset fusing eye-tracking and physiological data in VR comes from HP Labs (\cite{omnicept}) who conducted a study on large sample of 738 participants using a fusion of eye-tracking and physiological metrics and achieve up to 78\% accuracy. However, the only way they modulate the difficulty is through layering tasks on top of each other, arguing that managing three tasks is harder than two which is harder than a single task. Therefore, their high accuracy can be partly attributable to recognition of multiple tasks. It does not account, however, for difficulty modulations within each individual task. 

A recent study (\cite{clare2025}) claims a 'real-time' estimation of cognitive load, except that estimation was not conducted in real-time, but simulated after data collection. Similarly to \cite{omnicept} they use stacking of cognitive tasks instead of modulating difficulty of an individual task, risking multiple task recognition rather than actual cognitive load detection. As opposed to most studies, they report Leave-One-Participant-Out cross-validation (LOPO-CV) results, showing that using gaze and electrodermal activity (EDA) fusion, the best accuracy they could achieve was 69.59\% (F1 = 63.2\%) on a binary classification task. Moreover, this study was not done in VR.

To summarise, several studies attempt to estimate cognitive load using a variety of eye-tracking and physiological data modalities. They often use stacking of tasks as the only proxy for cognitive load manipulation, as opposed to manipulating the difficulty of the individual tasks. The machine learning models employed often rely on k-fold cross-validation which can lead to participant recognition from training and is not suitable for real-time applications which by their nature are leave-one-out paradigms as deploying to a new participant equals testing on a participant previously unseen in the data. The studies that do report leave-one-out results show large drops in accuracy (e.g. \cite{RN10}, \cite{clare2025}). Therefore, most of the results are inflated due to methodologies aimed at obtaining high accuracy scores in offline analysis rather than developing adaptive real-time systems. 

\subsection{Constraints of Current Offline Approaches for Cognitive Load Prediction}

Currently existing cognitive load estimation approaches rely predominantly on offline analysis of physiological data collected under highly controlled conditions, although the recent development of consumer-grade wearable devices has enormous potential to monitor daily workload objectively by acquiring physiological signals (\cite{anders2024unobtrusive}).

The constraints of the offline analysis further limit practical applicability. State-of-the-art methods in cognitive science either work offline only or involve bulky equipment hardly deployable in real-life settings (\cite{lagomarsino2022online}). This temporal disconnect between data collection and analysis further prevents the development of responsive, adaptive systems that could adjust to users' cognitive states in real-time.

\subsection{Adaptive Designs and Physiological Loops}

The domain of adaptive systems in Virtual Reality (VR) has been extensively reviewed by \cite{zahabi2020adaptive}, who established frameworks for training environments, and \cite{aranha2021adapting}, who focused on affective computing. Despite the breadth of research, the majority of existing systems rely on rule-based heuristics. These heuristics typically trigger adaptations based on performance metrics, such as error rates or reaction times (\cite{araujo2019}) or specific user behaviors, such as gaze duration. For instance, \cite{drey2020} utilized gaze tracking to generate adaptive hints; however, this adaptation was triggered by a simple timeout heuristic (looking at an object for too long) rather than a continuous estimation of cognitive state. Performance-based triggers are inherently reactive, modifying the system only after a user has already failed or stalled, rather than proactively managing load.

A recent systematic review by \cite{mortazavi2024dynamic} categorizes these performance-based approaches as indirect assessment methods, noting that while they are computationally efficient, they often fail to capture the internal cost of performance. Consequently, the field is witnessing a paradigm shift towards "data-driven" approaches utilizing Machine Learning (ML) and Deep Learning (DL). \cite{mortazavi2024dynamic} identify that while Reinforcement Learning (RL) and Neural Networks are becoming more prevalent for modeling complex player states, there remains a critical gap in applying these advanced architectures to real-time cognitive characteristics such as visual attention and working memory.

A significant portion of modern adaptive research focuses on affective states or contextual UI optimization rather than cognitive load. \cite{lindlbauer2019context} proposed an optimization-based approach for Mixed Reality interfaces, yet their system adapts UI placement based on context and task priority, not the user's mental resources. Similarly, recent advances in physiological computing often target emotion recognition. \cite{pinilla2023real} and \cite{quintero2025personalized} have demonstrated high accuracy in detecting affect using EEG and behavioral data. However, affect (valence and arousal) is distinct from cognitive load; an angry or happy user does not necessarily imply an overloaded one. 

Closer to the domain of workload, \cite{chiossi2023adapting} successfully utilised Electrodermal Activity (EDA) to adapt the VR environment. However, their adaptation strategy modulated visual complexity (e.g., reducing visual clutter) rather than the intrinsic difficulty of the task itself. While reducing extraneous load (visual clutter) is beneficial, it does not address the modulation of intrinsic cognitive load required for effective training progression. Furthermore, simple rule-based physiological loops, such as adjusting a meditative experience based on EDA thresholds (\cite{Salminen2024}), lack the temporal nuance to handle complex cognitive tasks. This limitation highlights the necessity for what \cite{mortazavi2024dynamic} describe as direct psychophysiological designs—systems that leverage sensors (e.g., EEG, Eye-tracking) to capture internal states. Crucially, the review notes that Long Short-Term Memory (LSTM) networks are particularly promising for this domain due to their ability to process the time-series nature of physiological data, yet their application in real-time VR settings remains underexplored.

\subsection{Advancing Real-Time Physiological Monitoring}

To address these gaps, we decided to leverage recent advances in machine learning, particularly deep learning architectures, offering promising solutions for real-time cognitive load estimation from physiological signals. Specifically, we propose a bidirectional LSTM architecture with an attention mechanism. Unlike traditional machine learning models, this architecture learns from the temporal relationships between inputs, making it well suited for tasks that involve sequential data. These architectural advances enable the development of models capable of processing continuous physiological streams and providing real-time feedback for adaptive systems.

\subsection{Study Rationale and Contributions}

To summarise, the key gaps in the research are a combination of the below: a) highly controlled task designs not suited for at-home use, b) over-reliance on within-subject cross-validation for cognitive load prediction using eye-tracking and/or physiological metrics leading to inflated offline results that would likely decrease when presented with a participant previously unseen in the data, c) the extensive use of task stacking for cognitive load manipulation risking the models being trained to distinguish number of tasks from eye-tracking patterns rather than actual cognitive load, d) existing adaptive studies relying on simple heuristics or, if using only physiological metrics, predicting visual complexity (extraneous load) or affect instead of intrinsic cognitive load. 

The present study addresses these critical gaps in the current understanding of real-time cognitive load estimation and adaptive VR training by introducing a novel approach that combines ecologically valid VR tasks with real-time physiological monitoring. Our research makes several key contributions. 
First, we present the first implementation of a real-time model-driven adaptive cognitive load estimation system deployed within an interactive VR environment using eye-tracking and physiological signals only.
Second, we evaluate the real-time model-driven design against a ground-truth real-time participant-driven design. 
Third, we compare a dual-task paradigm combining sustained attention with mental arithmetic to a single-task paradigm (sustained attention only) within a single study. This allows us to evaluate the generalisability of the model across different task designs.
Finally, our comprehensive evaluation framework, incorporating objective performance measures, subjective workload assessments, as well as physiological online monitoring, provides a holistic approach to understanding the complex relationships between cognitive load, adaptation strategies, and learning outcomes in VR environments. This multi-dimensional assessment approach addresses the lack of ecological validity that has hindered progress in adaptive cognitive training research.
With this investigation, we have advanced the field towards more effective, accessible, and scientifically grounded approaches to cognitive training with meaningful benefits to diverse populations whilst maintaining the necessary scientific rigor essential for clinical and educational applications.

\section{Methodology}
\begin{figure} 
    \centering
    \includegraphics[scale=0.45]
    {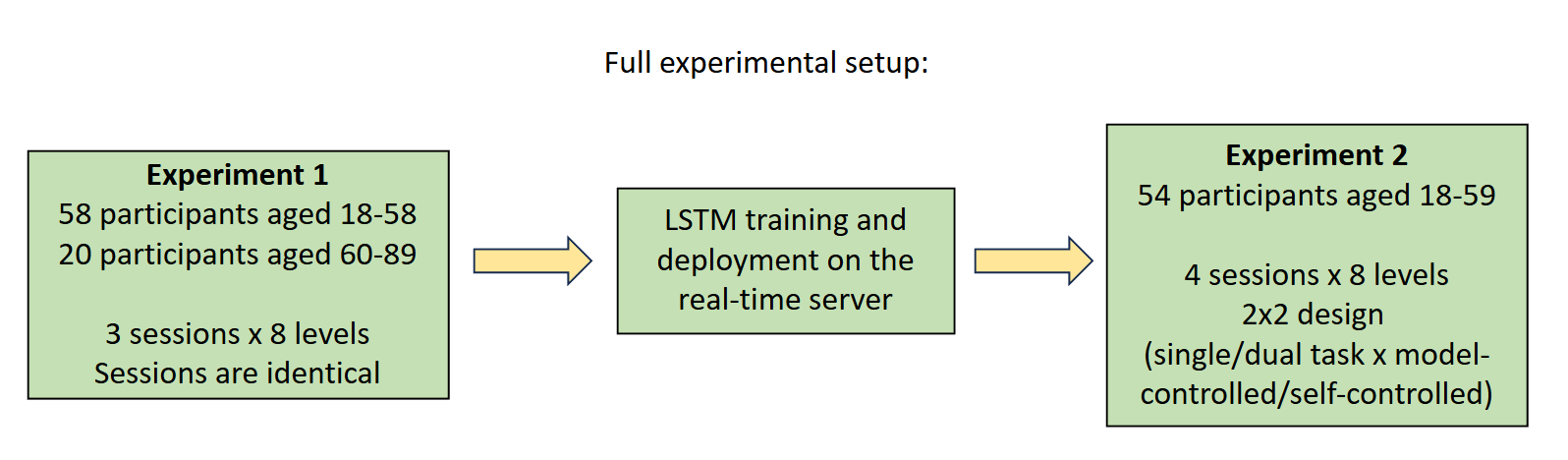}
    \caption{The flowchart depicting the key features and relationship of the experiments used in this study.}
    \label{fig:experiment_flow}
\end{figure}

The study consisted of two experiments - experiment 1 was used for data collection to train the model used later in real-time deployment in experiment 2. The relationship between the two experiments is presented in Figure \ref{fig:experiment_flow}.

\subsection{Participants}
\subsubsection{Demographics for experiment 1: training of the offline model (pre-deployment)}
We recruited participants in two stages for the same task. Firstly, 62 participants were recruited from the general working-age population aged 18-58 (M=29.94, SD=8.92). 32 were male and 30 female, 6 left-handed. These participants were recruited through online advertising and rewarded a £5 Amazon voucher for their participation. The second stage involved recruiting participants aged over 60. We recruited 20 participants aged 60-87 (M=71.7, SD=7.28). 6 of them were male, 14 female and all were right-handed. They were recruited from a University of the Third Age and compensated with £10 in cash. This amount was higher due to additional travel costs the participants needed to incur. We combined both samples for training of the offline (pre-deployment model). 4 participants from the general population group were excluded due to their experimental data not saving during the experiment due to a hard drive failure. Further 5 participants (4 from the general population and 1 from the older population) were excluded due to insufficient data quality. This resulted in a total sample of 73 participants aged 18-87 (M=41.26, SD=20.74). This included 30 male and 34 female participants, 6 of whom were left-handed.
\subsubsection{Demographics for experiment 2: real-time model evaluation}
We recruited 54 participants aged 18-59 (M = 28.5, SD = 8.4). 30 were female, 22 male and 2 non-binary. 3 of the participants were excluded due to issues with the real-time eye-tracking streaming, leading to invalid model predictions. The final sample included 51 participants aged 18-59 (M=28.6, SD=8.6), with 27 being female, 22 male, 2 non-binary. 49 of them were right-handed, 2 left-handed. Participants were recruited through online advertising.% and offered a £10 Amazon voucher for their participation.

\subsection{Experimental Design}
\subsubsection{Core experimental design for developing the model}
\begin{figure} 
    \centering
    \includegraphics[width=0.8\textwidth]{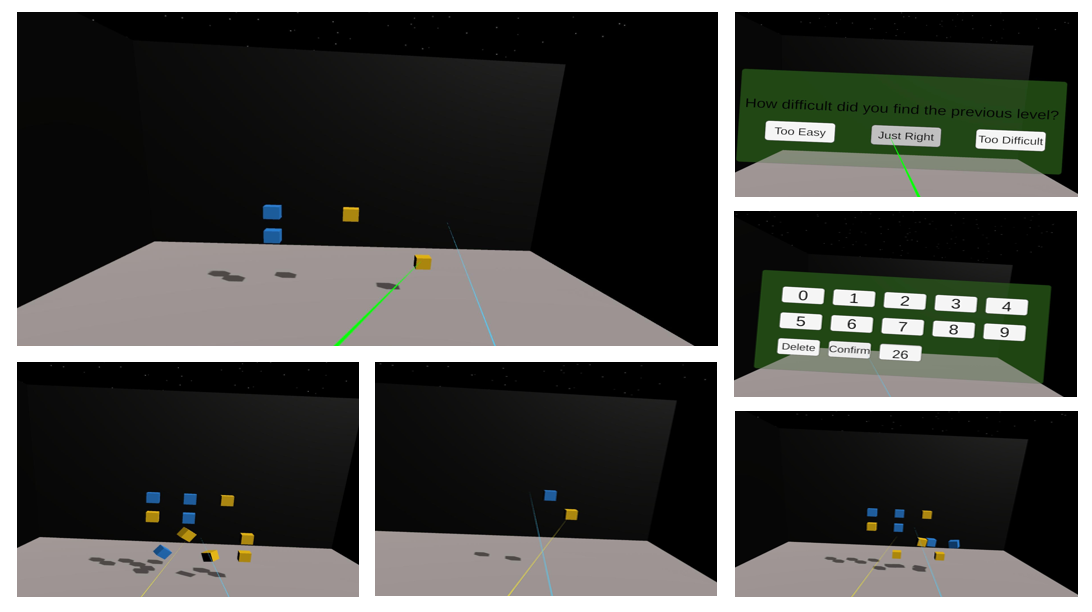}
    \caption{In-game view at different difficulty levels including mental arithmetic menu (middle right) and perceived difficulty selection menu (top right).}
    \label{fig:ingame}
\end{figure}

The experiment incorporated a novel sustained attention VR task (\cite{szczepaniak2024mldriven}), derived from continuous performance tasks (\cite{RN11}), in which participants were required to respond or withhold responses to stimuli across 8-minute periods of gameplay. Within the task, blue and yellow cubes appeared at random in the space before the participant. The participant held a controller in each hand, one projecting a blue pointer and the other a yellow pointer. The objective of the task was to eliminate the cubes materialising in front of the participant by employing the corresponding pointer (the blue pointer exclusively destroyed blue cubes, and the yellow pointer exclusively destroyed yellow cubes) (see Figure \ref{fig:ingame} for reference).
Four objective difficulty levels were established by regulating the intervals at which cubes appeared (with lower values indicating greater difficulty, as more cubes spawned within each interval) – 1.6s, 2.1s, 2.6s, 3.1s. The difficulty levels were determined through a pilot study involving 6 participants. To exemplify, at the most challenging difficulty level, a subsequent cube of the same color spawned 1.6s after the appearance of a blue or yellow cube.

The experiment consisted of three sessions, interspersed with 5-minute breaks. Each session was identical to each other as the aim was to increase the amount of data collected. During the breaks, participants completed a NASA-TLX questionnaire to evaluate their subjective cognitive load pertaining to the prior session. Within each session, every difficulty level was iterated twice, resulting in eight 60-second sub-sessions, separated by 10-second intervals.
The sequence of difficulty levels was randomized for each session.
The pivotal aspect of this design lies in its capability to adjust difficulty over brief intervals and solicit participant feedback without taking off the headset or the completion of an extensive questionnaire, thereby avoiding disruption to the experimental flow. This aspect is of particular importance for the subsequent real-time design.

\subsubsection{Framework for Real-time Model Adaptation}
\begin{figure} 
    \centering
    \includegraphics[scale=0.45]
    {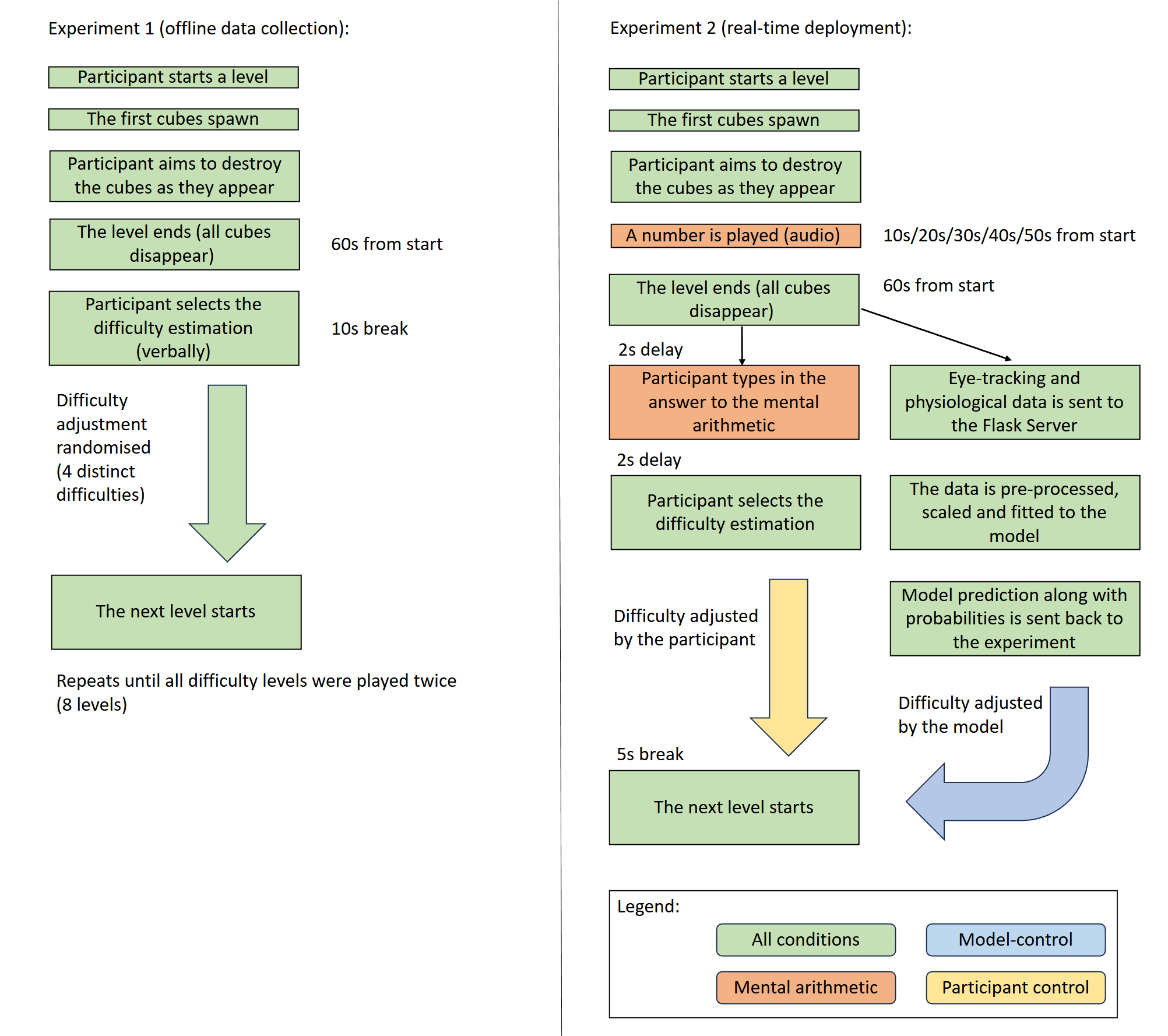}
    \caption{The flowchart depicting the experimental flow of the initial design used for data collection (experiment 1 on the left) and the adaptive design used to evaluate the real-time deployment of the model (experiment 2 on the right).}
    \label{fig:condition_designs}
\end{figure}

Following the first round of data collection aimed at developing a model that can be deployed in real-time, we decided to make some modifications to the experimental design. Firstly, the difficulty became adaptive instead of fixed. This means that participants always started at 2.6s delay coefficient (lower medium level in the offline design) and the difficulty was then guided by the model or participant, depending on condition. If the difficulty selection was \textit{Too Easy} the task would become more difficult by a factor of 0.3s. Conversely, if the difficulty selection was \textit{Too Difficult} the task would become easier by a factor of 0.3s. If the difficulty selection was \textit{Just Right}, it became harder by a factor of 0.1s to avoid stale difficulty throughout a session. Secondly, we added a dual task element to the design for two reasons: a) to allow for difficulty increase that is independent of physical dexterity, which was a risk at higher difficulty levels in the single task design, b) to test how well the model generalises to different types of tasks than the one it was trained on. Therefore, we added a simple mental arithmetic task, which involved 5 numbers between 1 and 9 being played (audio) throughout each 60s level at 10s intervals, starting 10s after level start. The participants were asked to provide the sum of the numbers at the end of each level using a keypad integrated into the VR experiment. This was followed by a menu presenting the difficulty choice (\textit{Too Easy, Just Right, Too Difficult}) (see Figure \ref{fig:ingame} for both menus). Following the difficulty choice, participants were given another 5s break before starting the next level. The participants played 8 levels in each condition, for a total of 4 conditions - two single task conditions, one controlled by the model, one by the participant, and two dual task conditions, one controlled by the model, one by the participant. The order in which the conditions were presented was randomised, however single task was always in spot 1 and 3 in the order and dual task in spot 2 and 4. This was to allow the participants to familiarise themselves with the experiment and to allow equal spacing between single/dual task conditions. The four random orders were: 1. model-guided single task, model-guided dual task, self-guided single task, self-guided dual task, 2. self-guided single task, self-guided dual task, model-guided single task, model-guided dual task, 3. self-guided single task, model-guided dual task, model-guided single task, self-guided dual task, 4. model-guided single-task, self-guided dual task, self-guided single task, model-guided dual task. For a visual breakdown of the 4 conditions see figure~\ref{fig:condition_designs}. The conditions were counter-balanced with 13 participants in all conditions except for condition 3 which had 15 participants. After exclusions order 1 had 13 participants, order 2 - 11 participants, order 3 - 15 participants and order 4 - 12 participants.
The rest of the task remained the same as in the first phase of the experiment (model development) and participants were asked to complete a NASA-TLX questionnaire at the end of each condition during the 5 minute breaks. 

\subsection{Materials}

The study was carried out utilizing an HTC VIVE PRO EYE headset featuring an integrated eye-tracker, which was linked to a desktop computer equipped with an NVIDIA RTX 2070 SUPER graphics card, AMD Ryzen 5 3600 processor, and 16GB of RAM.
Raw eye-tracking metrics were gathered via the HTC Vive Pro Eye headset at a 90 Hz frequency through a free Unity plugin. These encompassed eye openness, pupil diameter, sensor position, gaze origin and direction, and convergence distance. Physiological data, comprising photoplethysmography (PPG) and galvanic skin response (GSR), were acquired with a wrist-based Shimmer sensor at 51Hz and handled using the pyshimmer (\cite{PyS}) library in Python. A data collection pipeline was implemented to synchronize timestamps for eye-tracking data, physiological data, and in-task events via LabStreamingLayer (\cite{LSL}). Participants filled out a standard NASA-TLX questionnaire (\cite{hart1988development}) after each condition.

\subsection{Eye-tracking and Physiological Signal Processing}

\subsubsection{Eye-tracking Feature Extraction Pipeline}

Eye-tracking data was processed using a comprehensive pipeline designed to extract robust gaze behavior and pupil dynamics metrics. Raw eye-tracking data was collected at 90 Hz from the integrated HTC Vive Pro Eye tracker and processed through several stages following established protocols for VR-based cognitive load assessment (\cite{larsen2024method}).

\textbf{Preprocessing and Noise Reduction}: Raw gaze direction vectors were computed by averaging left and right eye 3D gaze directions and normalizing to unit vectors. Direction signals were smoothed using Savitzky-Golay filtering (window length = 11, polynomial order = 2) to reduce high-frequency noise while preserving important signal characteristics. This approach maintains peak information integrity while filtering signal artifacts, following established practices for optimal noise reduction without signal distortion in eye-tracking applications (\cite{raju2021filtering}).

\textbf{Blink Detection and Interpolation}: Blinks were detected using adaptive thresholding based on pupil diameter fluctuations, with eye openness values below 70\% classified as blinks. Rolling statistics (100ms windows, \~ 9 samples) were used to establish dynamic thresholds that account for individual variations in eye anatomy and blink patterns (\cite{nystrom2024blink}). Blink segments were extended by 50ms margins on each side to account for partial blinks, and missing data during blinks was interpolated using linear interpolation following the Identification-Artifact Correction (I-AC) protocol.

\textbf{Velocity and Acceleration Computation}: Angular velocity was calculated from consecutive gaze direction samples using the formula: $velocity = \sqrt{(\Delta x)^2 + (\Delta y)^2} \times \text{sampling\_rate}$
, where $\Delta$x and $\Delta$y represent frame-to-frame changes in normalized gaze direction. Acceleration was computed as the temporal derivative of velocity, providing input for advanced saccade characterization.

\textbf{Fixation and Saccade Classification}: A Hidden Markov Model (HMM) with two Gaussian states was employed to classify each sample as either fixation (state 0) or saccade (state 1) (\cite{kim2020hidden}). The HMM was fitted using velocity profiles with 2D Gaussian emission densities, providing more robust classification than simple velocity thresholding methods. This approach has been shown to achieve superior performance in identifying saccade patterns compared to traditional threshold-based methods (\cite{coutrot2018scanpath}) and is of particular importance in real-time designs where all the calculations need to be performed during the experiment.

\textbf{Advanced Saccade Metrics}: Beyond basic saccade counting, advanced metrics were extracted including peak velocity, mean acceleration/deceleration phases, velocity variability, and acceleration-deceleration ratios (\cite{distasi2010saccadic}). Peak saccadic velocity was extracted as the highest velocity during individual saccades, with main sequence relationship normalization for amplitude effects. These metrics provide insights into saccade dynamics that correlate with cognitive load and demonstrate sensitivity sufficient to distinguish among cognitive demand levels (\cite{bachurina2022multiple}).

\textbf{Pupil Dynamics Analysis}: Pupil diameter was processed to extract both static measures (mean, standard deviation, range) and dynamic measures including linear trend (slope), constriction velocity (rate of pupil size decrease), and dilation velocity (rate of pupil size increase) with a 2nd degree Butterworth filter applied for preprocessing (\cite{krejtz2018eye}).

\subsubsection{Physiological Signal Processing}

\textbf{Photoplethysmography (PPG) Processing}: PPG signals were collected at 51 Hz using a wrist-mounted Shimmer sensor and processed using the HeartPy library (\cite{van2019heartpy}). Signal preprocessing included 2nd-order Butterworth low-pass filtering (cutoff = 3 Hz, order = 3) to minimize PPG waveform distortion (\cite{lapitan2024estimation}), followed by peak enhancement using two iterations of linear transformation. Heart rate variability metrics were computed using segmentwise analysis to extract beat-per-minute (BPM), standard deviation of normal-to-normal intervals (SDNN), root mean square of successive differences (RMSSD) and percentage of successive RR intervals differing by more than 50ms (pNN50) (\cite{solhjoo2019heart}).

\textbf{Galvanic Skin Response (GSR) Processing}: GSR signals were processed to extract three primary metrics: mean conductance level, conductance variability (standard deviation), and linear trend (slope) (\cite{zangroniz2017observing}). Missing values were handled using forward-fill followed by backward-fill interpolation to maintain signal continuity. Baseline-relative averaging was applied following established protocols for VR stress response measurement.

\subsubsection{Temporal Windowing and Feature Aggregation} 

To capture temporal dynamics while maintaining computational efficiency for real-time processing, data from each 60-second experimental level was divided into four sequential 15-second windows (1,325 raw samples per window at 90 Hz for eye-tracking and 765 samples per window for physiological data at 51Hz) (\cite{zanetti2022real}). This windowing approach enables the model to learn temporal patterns in cognitive load changes throughout each task level while providing sufficient temporal resolution for real-time applications (\cite{sarailoo2022assessment}).

Windowing parameters were selected based on established research showing 15-second windows provide optimal balance between temporal resolution and classification accuracy for cognitive load assessment as smaller windows have been shown to better discriminate between cognitive states compared to longer analysis periods (\cite{behrens2021quantifying}).

For each window, 21 eye-tracking features and 7 physiological features were extracted, resulting in 28 features per window. The complete feature set included:

\textbf{Eye-tracking features per window}: blink count, blink rate, fixation count, mean/std fixation duration, saccade count, mean/std saccade amplitude, saccade-fixation ratio, advanced saccade metrics (peak velocity, acceleration/deceleration characteristics), and comprehensive pupil measures (mean, std, range, slope, constriction/dilation velocities).

\textbf{Physiological features per window}: GSR metrics (mean, std, slope) and heart rate variability metrics (BPM, SDNN, RMSSD, PNN50).

\subsection{Real-time Multi-modal Cognitive Load Estimation Model}
\begin{figure} 
    \centering
    \includegraphics[width=0.8\textwidth]{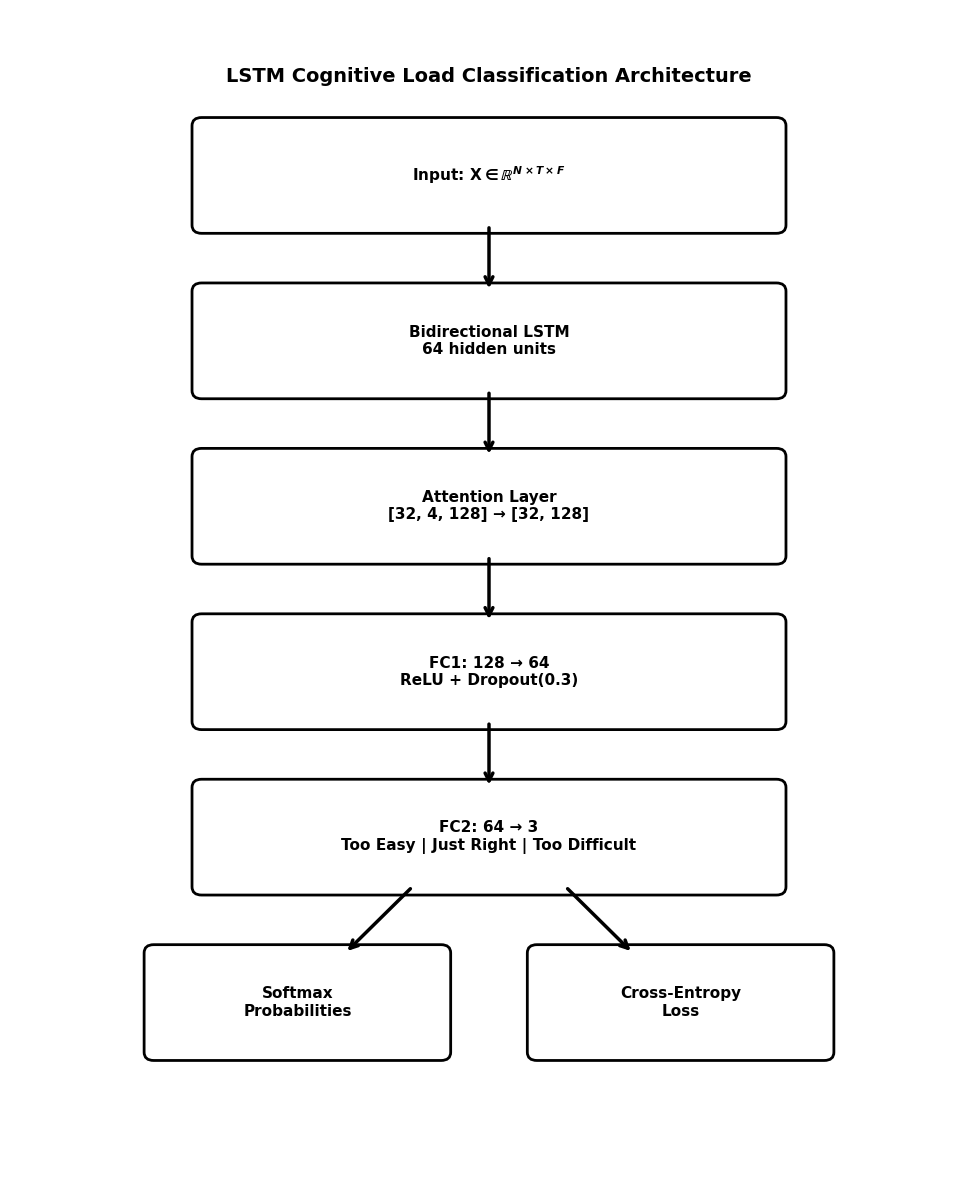}
    \caption{Diagram of the LSTM classifier architecture.}
    \label{fig:lstm_classifier}
\end{figure}

\subsubsection{Attention-Enhanced Bidirectional LSTM Architecture}

The cognitive load estimation model employed a bidirectional Long Short-Term Memory (LSTM) architecture with self-attention mechanisms, specifically designed to capture temporal dependencies in physiological and eye-tracking signals (\cite{pham2021timefrequency}), see Figure \ref{fig:lstm_classifier}. The model architecture consisted of several key components optimized for cognitive load classification from multimodal physiological data.

\textbf{Input Layer Configuration}: The model processed input tensors $\mathbf{X} \in \mathbb{R}^{N \times T \times F}$, where $N$ represents the batch size, $T=4$ denotes the temporal sequence length, and $F=28$ represents the dimensionality of the concatenated feature vector per time step. 
The feature space comprised 21 oculometric parameters and 7 physiological markers extracted through standardized preprocessing pipelines to ensure temporal alignment and feature normalization across modalities.
%sequences of shape [batch\_size, 4, 28], where 4 represents the temporal windows and 28 represents the concatenated eye-tracking and physiological features per window.

\textbf{Bidirectional LSTM Processing}: A bidirectional LSTM with 64 hidden units processed the input sequences, following state-of-the-art work on physiological signal classification (\cite{asgher2020enhanced}). Bidirectionality enabled the model to capture both forward and backward temporal dependencies, crucial for understanding cognitive load patterns that may manifest with temporal delays or anticipatory effects. The network used hyperbolic tangent for state activation and sigmoid for gate activation.

\textbf{Attention Mechanism}: A self-attention layer was implemented to automatically weight the importance of different temporal segments \cite{zhang2022mental}. The attention mechanism computed contextual importance through:
\begin{equation}
    \label{eq:attention1}
    \text{attention\_weights} = \text{softmax}(W_a \cdot h_t + b_a)
\end{equation}

\begin{equation}
    \label{eq:attention2}
    \text{context\_vector} = \sum_{t} (\text{attention\_weights}_t \odot h_t)
\end{equation}

where $h_t$ represents the LSTM hidden state at time $t$, $W_a$, $b_a$ are learnable attention parameters, $\cdot$ denotes matrix-wise multiplication and $\odot$ denotes element-wise multiplication. This self-attention mechanism enabled the model to focus on the most informative segments while maintaining comprehensive feature integration across the entire sequence of eye-tracking and physiological data.

\textbf{Classification Head}: The final classification layer consisted of a two-layer fully connected network with ReLU activation and dropout regularization (0.3). The network mapped the attention-weighted context vector to class probabilities for three difficulty categories: "Too Easy," "Just Right," and "Too Difficult" based on a softmax activation.

\textbf{Class Imbalance Mitigation}: To address class imbalance in cognitive load datasets, a multi-stage augmentation pipeline was implemented. Initial class balancing was achieved through participant-wise median target count normalisation, followed by synthetic sample generation. 
The augmentation pipeline comprised three complementary strategies: (1) \textbf{temporal jittering} with Gaussian noise injection ($\sigma$ = 0.05) to improve model robustness to temporal variations. (2) \textbf{Intra-class window mixing} where temporal segments from different sequences within the same 'difficulty' class were combined (e.g., concatenating windows 1-2 from participant A with windows 3-4 from participant B, both labeled "Low" difficulty), (3) \textbf{cross-class interpolation} using weighted linear combinations of "Low" and "High" difficulty samples to generate synthetic "Medium" difficulty sequences with interpolated weight of $\alpha \in [0.45-0.55]$. 
To improve the classification of 'medium' level difficult, we employed \textbf{structured temporal pattern synthesis} using predefined arrangements:
$(Medium_1 \to Low_3 \to High_2 \to Medium_4 \text{ and } Medium_4 \to High_3 \to Low_2 \to Medium_1)$, where subscripts indicate temporal position within the original sequence. This approach captured transitional cognitive load states characteristic of intermediate difficulty levels. 
%, and (3) pattern-based synthesis for medium difficulty using specific temporal arrangements (Medium1\_Low3\_High2\_Medium4 and Medium4\_High3\_Low2\_Medium1 patterns) to capture transitional cognitive load states, where the numbers indicate the position in the sequence of the original segment.

\textbf{Dynamic Class Weighting}: Class weights were computed based on the inverse frequency scaling with manual adjustments to enhance edge case detection. Applied weights were 1.5× for \textit{Too Easy}, 1.5× for \textit{Just Right}, and 1.7× for \textit{Too Difficult} categories, multiplied by their respective inverse frequency weights to address challenging classification boundaries (\cite{ahmad2023framework}).

\subsection{Model Training Configuration}

Key training parameters included:
\begin{itemize}
    \item \textbf{Loss Function}: Weighted categorical cross-entropy for standard classification
    \item \textbf{Optimizer}: Adam optimizer (learning rate = 0.001, weight decay = 0.01)
    \item \textbf{Regularization}: Dropout (0.3) and early stopping (patience = 15 epochs) to prevent overfitting
    \item \textbf{Batch Size}: 32, optimized for memory constraints and convergence stability
    \item \textbf{Maximum Epochs}: 100 with early stopping criteria based on validation accuracy and loss
    \item \textbf{Network Architecture}: Single bidirectional LSTM layer with 64 hidden units
    \item \textbf{Data Augmentation}: Multi-strategy approach including Gaussian jitter ($\sigma$ = 0.05), temporal window mixing, and cross-class interpolation generating up to 380 additional synthetic samples per class
    \item \textbf{Feature Preprocessing}: Standard normalization (z-score) applied to all input features fitted on training data only
    \item \textbf{Validation Split}: 10\% stratified split from training data for model selection and early stopping
    \item \textbf{Training Strategy}: Edge case prioritization with weighted loss to improve detection of "Too Easy" and "Too Difficult" classifications over "Just Right" categories
\end{itemize}

\subsection{Cross-Validation, Generalisability \& Interpretability}

\textbf{Leave-One-Participant-Out Protocol}: The model was trained using data from 74 participants collected in the initial model development phase. Training employed leave-one-participant-out (LOPO) cross-validation to assess generalization across participants (\cite{kunjan2021necessity}). This validation approach is essential for physiological signal classification as it prevents data leakage between training and testing sets while providing realistic estimates of cross-participant generalization performance (\cite{wang2023characterisation}).

\textbf{Nested Validation Protocol}: Within each LOPO fold, the training set (73 participants) was further partitioned using stratified 90/10 training-validation splits to enable unbiased hyperparameter selection and early stopping. 
This nested validation approach ensured unbiased model selection within each fold while providing robust generalization estimates across the participant population.

\subsubsection{Feature Scaling and Normalization} 

All features were standardized using a global z-score normalization strategy, where scaling parameters were derived from the training set. These fixed parameters were applied directly to the real-time data stream during deployment, treating the training distribution as representative of the target population. This approach was selected to facilitate a zero-calibration framework, enabling the system to generate immediate, one-shot predictions for novel users without the requirement for obtaining pre-session baselines or participant-specific adaptation.

\subsubsection{Feature importance analysis}

To interpret the key decision markers behind our model, we used SHAP (SHapley Additive exPlanations) (\cite{lundberg2017unified}), which is a versatile approach used to interpret the results of machine learning models. It calculates the contribution of each feature to the prediction of a specific instance, regardless of the model used. This method leverages Shapley values from cooperative game theory to distribute the overall value generated by a group of players fairly among individual players. By assigning SHAP values to each feature, one can effectively rank the features based on their importance in the prediction process (\cite{lundberg2017unified}). As opposed to traditional explainability approaches which estimate feature importance by shuffling one feature at a time, SHAP is relatively resilient to correlated variables.

\subsection{Real-time Deployment for VR integration}

The trained cognitive load estimation model was deployed using a distributed microservice architecture built on Flask, enabling streaming and processing of eye-tracking and physiological data in real-time (\cite{ahmad2023framework}). The deployment pipeline achieved sub 500ms end-to-end latency, from data acquisition to cognitive load estimation, suitable for responsive adaptive VR applications.

\textbf{Data Stream Management and Synchronization}: Multimodal data acquisition was orchestrated through  LabStreamingLayer (LSL) protocol \cite{LSL}, which handled temporal synchronization between heterogeneous sensor streams operating at different sampling rates. 

\textbf{Real-time Feature Extraction Pipeline}: The preprocessing and feature extraction methods used during training were applied to streaming data, maintaining consistency between training and deployment conditions. This includes identical filtering parameters, windowing strategies, and normalization procedures.

\textbf{Model Inference Engine}: PyTorch models were loaded with pre-trained weights and executed on incoming feature sequences to generate real-time predictions with class probabilities. The system maintained model state across inference calls to ensure temporal continuity.

\textbf{Adaptive VR Integration}: Model predictions were integrated with the VR task difficulty adjustment system, enabling automatic difficulty modulation based on estimated cognitive load levels.

This real-time architecture enabled cognitive load estimation with minimal latency while maintaining the robustness necessary for continuous operation in interactive VR environments.

\section{Results}
\subsection{Model Evaluation (pre-deployment)}

Model evaluation was assessed through comprehensive evaluation on held-out test data using participant-wise analysis with aggregated cross-validation results. The evaluation protocol employed leave-one-participant-out cross-validation to ensure robust generalization estimates across the participant population while preventing data leakage between training and testing phases. 

\begin{figure}
    \centering
    \includegraphics[width=0.8\textwidth]{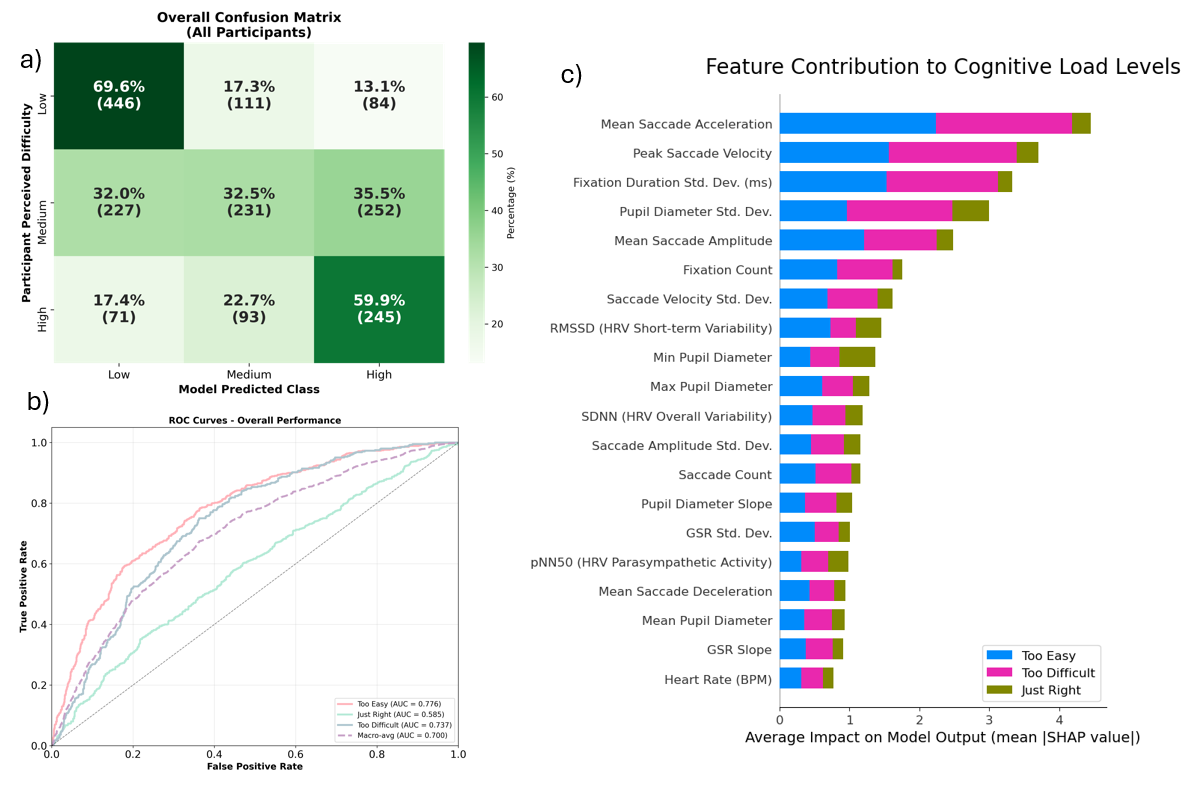}
    \caption{Confusion matrix, ROC curves and feature importance rating for the offline (pre-deployment) model.}
    \label{fig:methodology_evaluation}
\end{figure}

\textbf{Classification Performance:}
Detailed performance metrics are summarized in Table \ref{tab:overall_metrics} and Figure  \ref{fig:methodology_evaluation}. Table \ref{tab:overall_metrics} includes accuracy, precision, recall, AUC-ROC, AUPRC and F1-scores.  Figure \ref{fig:methodology_evaluation} presents comprehensive performance metrics including a confusion matrix (Figure \ref{fig:methodology_evaluation}a) showing classification accuracy across difficulty levels and ROC curves (Figure \ref{fig:methodology_evaluation}b) demonstrating discriminative power for each class with corresponding area-under-curve values.  
The model demonstrated strong discriminative performance across cognitive load categories, with results aligning with our strategic emphasis on edge-case detection over \textit{Just Right} classifications. This design priority enables the adaptive system, developed subsequently, to identify cognitive states requiring immediate difficulty adjustment while tolerating minor misclassifications within the optimal engagement zone. Critically, the model exhibited a low rate of severe misclassifications (8.81\% of all predictions) where \textit{Too Difficult} samples were incorrectly predicted as \textit{Too Easy} or vice versa - errors that would result in inappropriate difficulty adjustments and potential user frustration or disengagement.
These results demonstrate the model's readiness for real-time deployment in adaptive VR environments while maintaining acceptable performance across diverse user populations. 

\textbf{Feature Importance Analysis:}
The SHAP analysis of feature importance revealed that oculometric measures dominated model predictions, with the top 7 contributing features derived exclusively from eye-tracking data (Figure \ref{fig:methodology_evaluation}c). The most informative features were primarily saccadic metrics (specifically mean acceleration, peak velocity, and mean amplitude) and fixation-related parameters (standard deviation of duration and count), alongside pupil diameter variability. Physiological measures played a secondary role, with short-term Heart Rate Variability (RMSSD) appearing as the eighth most important feature, while blink frequency patterns were notably absent from the top contributions.

\begin{figure}
  \centering
  \includegraphics[width=1\textwidth]{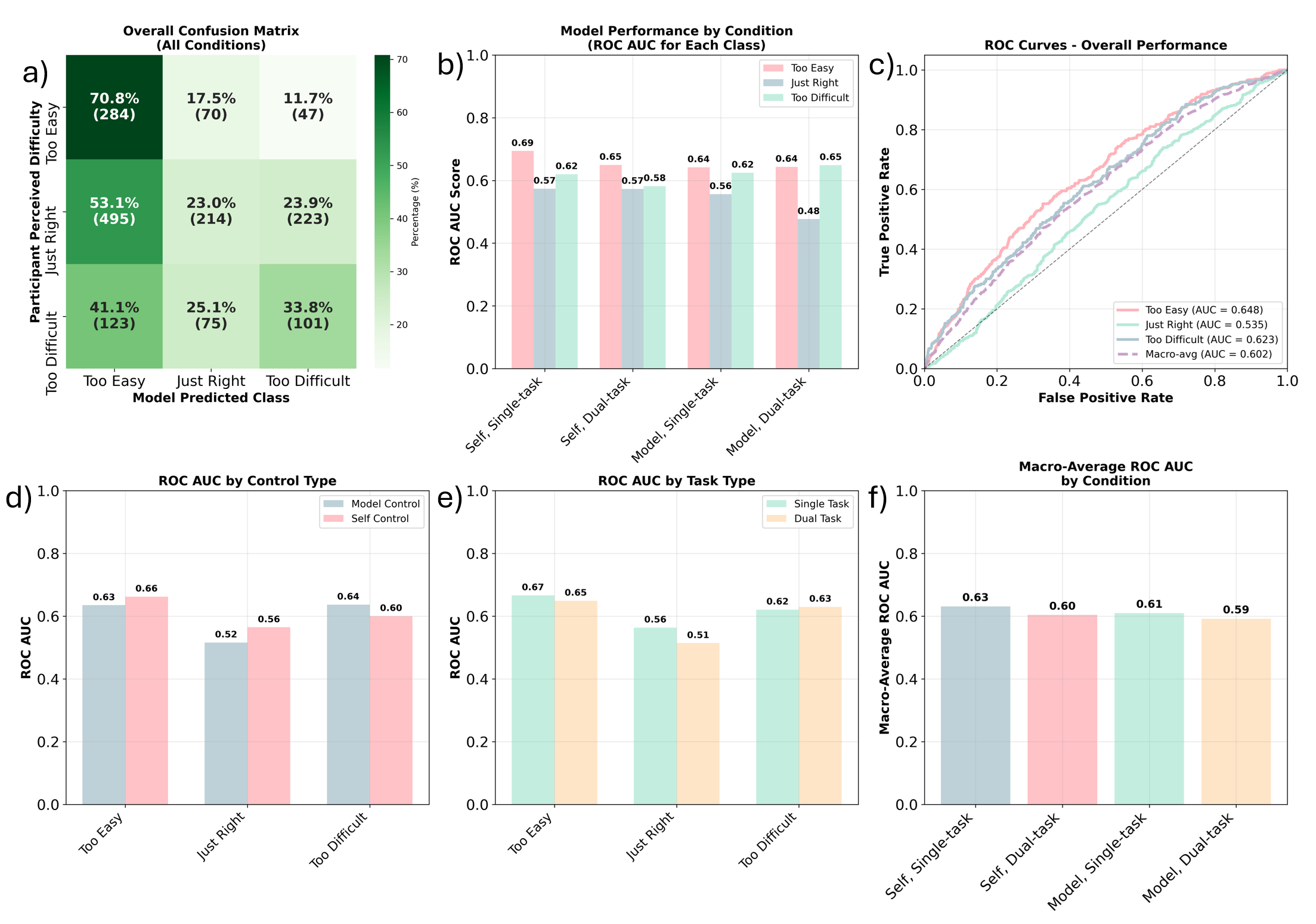}
  \caption{Complete model evaluation across all experimental conditions. (a) Overall confusion matrix showing prediction accuracy with percentages and counts, (b) Model performance by condition showing ROC AUC scores for each difficulty class, (c) ROC curves for overall performance, (d) ROC AUC comparison by control type, (e) ROC AUC comparison by task type, (f) Macro-average ROC AUC by individual condition.}
  \label{fig:model_evaluation}
\end{figure}

\subsection{Evaluation of Physiologically-Informed Adaptive Difficulty Prediction in Virtual Reality (Analysis of the Classifier Performance During Deployment)}
\begin{table}

\caption{Overall Metrics Averaged Across LOPO-CV Folds (pre-deployment)}
\centering
\begin{tabular}{lcccc}
\toprule
\textbf{Metric} & \textbf{Mean} & \textbf{Std} & \textbf{Min} & \textbf{Max} \\
\midrule
Accuracy & 52.3\% & 14.5\% & 16.7\% & 83.3\% \\

F1-Score (Macro-Averaged) & 49.7\% & 16.2\% & 13.7\% & 80.8\% \\
 
Precision (Macro-Averaged) & 58.9\% & 17.4\% & 15.2\% & 87.5\% \\

Recall (Macro-Averaged) & 54.0\% & 15\% & 20\% & 90\% \\

AUC-ROC (Macro-Averaged) & 73.7\% & 10.7\% & 47\% & 98\% \\

AUPRC (Macro-Averaged) & 62.5\% & 12.3\% & 35.8\% & 97.6\% \\

\bottomrule
\end{tabular}

\label{tab:overall_metrics}
\end{table}

\begin{table} 
\caption{Overall Metrics Averaged Across All Real-Time Predictions (During Deployment)}

\centering
\begin{tabular}{lcccc}
\toprule
\textbf{Metric} & \textbf{Mean} & \textbf{Std} & \textbf{Min} & \textbf{Max} \\
\midrule
Accuracy & 36.7\% & 11.9\% & 6.2\% & 59.4\% \\
F1-Score (Macro-Averaged) & 32.6\% & 10.0\% & 8.0\% & 56.0\% \\
Precision (Macro-Averaged) & 43.2\% & 11.3\% & 13.7\% & 72.0\% \\
Recall (Macro-Averaged) & 42.1\% & 12.5\% & 4.6\% & 65.7\% \\
AUC-ROC (Macro-Averaged) & 63.0\% & 10.7\% & 41.3\% & 82.0\% \\
AUPRC (Macro-Averaged) & 49.5\% & 11.6\% & 31.4\% & 80.4\% \\

\bottomrule
\end{tabular}
\label{tab:overall_metrics_rt}
\end{table}

Real-time classification performance was assessed through comprehensive analysis of model predictions against participant-reported difficulty ratings across all experimental conditions.
The evaluation leveraged continuous model predictions throughout the experimental protocol, enabling comparison of 32 trials per participant against ground-truth subjective difficulty assessments, yielding 1,632 (32x51) total prediction-rating pairs for analysis.

\subsubsection{Overall Classification Performance}
The deployed model achieved 36.7\% overall accuracy in predicting participant difficulty ratings during real-time operation, representing a substantial performance degradation from the 52.3\% offline training accuracy. This performance gap highlights the inherent challenges of translating laboratory-trained models to dynamic, interactive environments where temporal context, user adaptation, and real-time processing constraints impact classification reliability.
The confusion matrix (Figure~\ref{fig:model_evaluation}a) revealed systematic bias toward "Too Easy" predictions, with the most common error being classification of \textit{Just Right} trials incorrectly as \textit{Too Easy} (30.3\% of all trials). 
Class-specific precision demonstrated the anticipated training strategy effects: moderate precision for \textit{Just Right} classifications (0.60), reflecting the model's design emphasis on edge-case detection, while \textit{Too Easy} (0.31) and \textit{Too Difficult} (0.27) categories showed lower precision values.

Model prediction accuracy varied significantly across experimental conditions (Figure \ref{fig:model_evaluation}b).
Comparison between control types (Figure~\ref{fig:model_evaluation}d) and task types (Figure~\ref{fig:model_evaluation}e) confirmed the accuracy findings, with self-control conditions showing marginally better AUC values (0.610 vs. 0.597) and single task conditions outperforming dual task conditions (0.618 vs. 0.599). Individual condition analysis (Figure~\ref{fig:model_evaluation}f) confirmed that Self Single Task achieved the highest macro-average AUC (0.631), while Model Dual Task showed the lowest performance (0.592).
The model's design priority to detect edge cases (\textit{Too Easy} and \textit{Too Difficult}) over \textit{Just Right} classifications is evident in these results, aligning with the intended training objective.

\subsection{Subjective Difficulty Analysis}
\begin{figure} 
    \centering
    \includegraphics[width=1\linewidth]{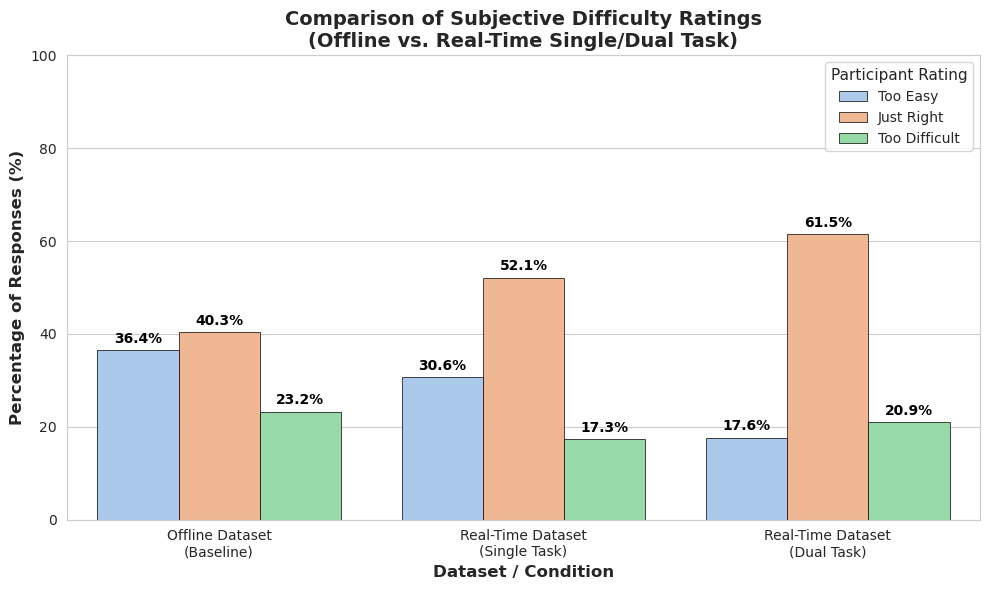}
    \caption{Distribution of Subjective Difficulty Ratings. The chart compares the original dataset (baseline) with the current study's Single and Dual Task conditions.}
    \label{fig:subjective_ratings}
\end{figure}

To validate the efficacy of the experimental manipulation, we analyzed the distribution of subjective difficulty ratings (\textit{Too Easy, Just Right, Too Difficult}) across conditions. Figure \ref{fig:subjective_ratings} compares the rating distributions of the original baseline dataset against the Single and Dual Task conditions of the current study.

\subsubsection{Perceived Difficulty Comparison Between Single and Dual Real-Time Designs}
The introduction of the secondary auditory task successfully induced a shift in perceived difficulty. In the Single Task condition, 30.6\% of segments were rated as \textit{Too Easy}, whereas this dropped significantly to 17.6\% in the Dual Task condition ($\chi^2 = 158.70, p < 0.001$). Conversely, reports of \textit{Too Difficult} ratings rose from 17.3\% to 20.9\% ($\chi^2 = 14.65, p < 0.001$). An omnibus Chi-square test confirmed that the overall distributional shift was statistically significant ($\chi^2 = 159.58, p < 0.001$), indicating that the dual-task protocol effectively increased cognitive load as perceived by participants.

\subsubsection{Comparison of the Real-Time and Offline Datasets}
We also compared the real-time Single Task condition to the offline dataset. The offline dataset showed a more polarized distribution compared to the real-time Single Task, where responses converged more heavily toward the \textit{Just Right} category. Specifically, the real-time Single Task condition had significantly fewer \textit{Too Easy} ratings ($\chi^2 = 34.18, p < 0.001$) and fewer \textit{Too Difficult} ratings ($\chi^2 = 48.29, p < 0.001$) compared to the offline baseline. 

\subsection{Game Performance Analysis: Influences of Control Mechanism and Task Complexity}
\begin{figure} 
    \centering
    \includegraphics[width=1\textwidth]{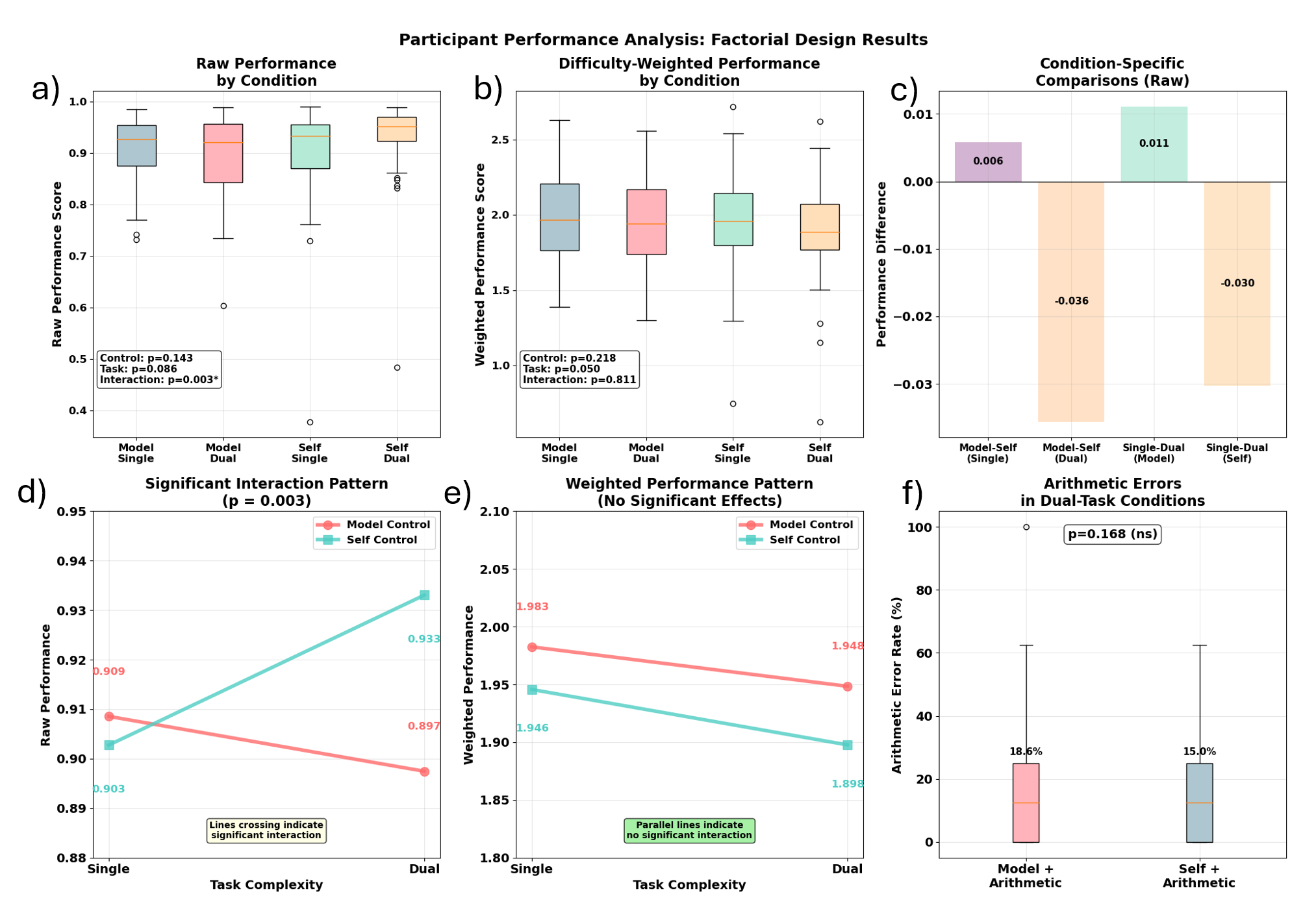}
    \caption{Integrated performance and error analysis across experimental conditions based on 2×2 repeated measures ANOVA. (a) Raw game performance scores by condition showing significant Control × Task Complexity interaction (p = 0.003) with equivalent model-self game performance in single tasks but self-control advantage in dual tasks; (b) Difficulty-weighted game performance revealing no significant main effects or interactions (all p > 0.05), indicating equivalent adaptive system effectiveness across conditions; (c) Condition-specific performance comparisons highlighting the interaction pattern, showing how model-self differences vary by task complexity; (d) Raw game performance interaction plot demonstrating significant crossing pattern (p = 0.003) where control type effectiveness depends on task complexity; (e) Weighted game performance interaction plot showing parallel lines indicating no significant interaction (p = 0.811), confirming equivalent system performance; (f) Arithmetic error rates in dual-task conditions showing numerically higher errors for model-control (18.6\%) versus self-control (15.0\%, p = 0.168, ns). Statistical analyses used repeated measures ANOVA for factorial comparisons and paired t-tests for specific contrasts. Box plots show median, quartiles, and outliers; interaction plots display condition means with statistical annotations.}
    \label{fig:integrated_performance}
\end{figure}

To evaluate how control type (model-controlled vs. participant-controlled) and task complexity (single-task vs. dual-task) affect game performance in adaptive virtual reality environments, we conducted a 2×2 repeated measures ANOVA. In this context, game performance refers to the participant's ability to complete in-game objectives under varying difficulty conditions. This statistical test was designed to identify: (1) whether the control mechanism influences performance outcomes, (2) whether task complexity alters performance, and (3) whether there is an interaction effect between these two factors. For example, whether the impact of control type depends on task complexity.

Measuring game performance in adaptive virtual reality tasks is a nuanced problem, particularly when task difficulty varies dynamically based on participant ability. Here we use three complementary metrics of game performance to obtain a robust view of performance:
\begin{itemize}
    \item \textbf{Raw game performance:} proportion of cubes destroyed out of total cubes spawned (range 0-1; higher is better). This can be used to illustrate the bias of participants in judging their performance but is not an accurate representation of actual game performance as it does not take into account the difficulty at which a given score was achieved. 
    \item \textbf{Difficulty-weighted game performance:} raw score multiplied by a difficulty modifier (range 0-4.3; higher is better). This is the most accurate representation of participant game performance as it provides a nuanced score taking into account both the difficulty and proportion of cubes destroyed. In other words, destroying a high proportion of cubes at a high difficulty level is objectively a better performance than destroying the same amount of cubes at a lower difficulty level. 
    \item \textbf{Arithmetic error rate:} number of incorrect responses in dual-task conditions (the less the better)
\end{itemize}  

A series of 2×2 repeated measures ANOVAs was conducted with Control Type (Model vs Self) and Task Complexity (Single vs Dual) as within-subjects factors and raw game performance as well as difficulty-weighted game performance as dependent variables. A paired t-test was used to determine whether there was a difference in arithmetic errors in the dual conditions depending on Control Type. 
Prior to analysis, assumption checks were performed for the primary dependent variables. Outlier detection utilizing the Interquartile Range (IQR) method identified a limited number of outliers across conditions. Shapiro-Wilk tests indicated that raw game performance data violated the assumption of normality ($p < .001$) due to ceiling effect expected in levels that were too easy and did not sufficiently challenge participants. To ensure statistical rigor, we conducted the primary analysis using standard repeated measures ANOVA and validated significant findings using a robustness check with arcsine square-root transformed data.
The key results for this section are visualised in Figure \ref{fig:integrated_performance}.
Analysis of raw game performance revealed a significant Control Type $\times$ Task Complexity interaction ($F(1, 50) = 7.16, p = .008$). While performance was statistically equivalent in single-task conditions ($M_{Model} = 0.91$ vs. $M_{Self} = 0.90$), participants in the self-control condition achieved significantly higher raw accuracy in dual-task conditions ($M_{Self} = 0.93$) compared to the model-control condition ($M_{Model} = 0.90$). This interaction effect was confirmed by the robustness check on transformed data ($F(1, 50) = 11.44, p = .001$), indicating that the divergence in strategies is a robust phenomenon and not an artifact of the data distribution.

Critically, difficulty-weighted game performance analysis revealed a significant main effect for Task Complexity ($F(1,50) = 4.045, p = 0.050$), indicating that game performance differed between single and dual tasks when accounting for task difficulty, with no significant effects for Control Type ($F(1, 50) = 1.55, p = .219$) or the Control × Task Complexity interaction ($F(1, 50) = 4.04, p = .050$). This suggests that while the adaptive system maintained equivalent challenge calibration across control types, task complexity influenced game performance outcomes when weighted by difficulty.

Mental arithmetic errors showed no significant differences ($t(50) = 1.400, p = 0.168, d = 0.20$) for Control Type. The absence of significant differences in weighted game performance as well as arithmetic errors confirms that the adaptive system successfully maintained appropriate challenge levels.

\subsection{Adaptive Control Effectiveness: Difficulty Trends and Convergence}

\begin{figure} 
  \centering
  \includegraphics[width=1\textwidth]{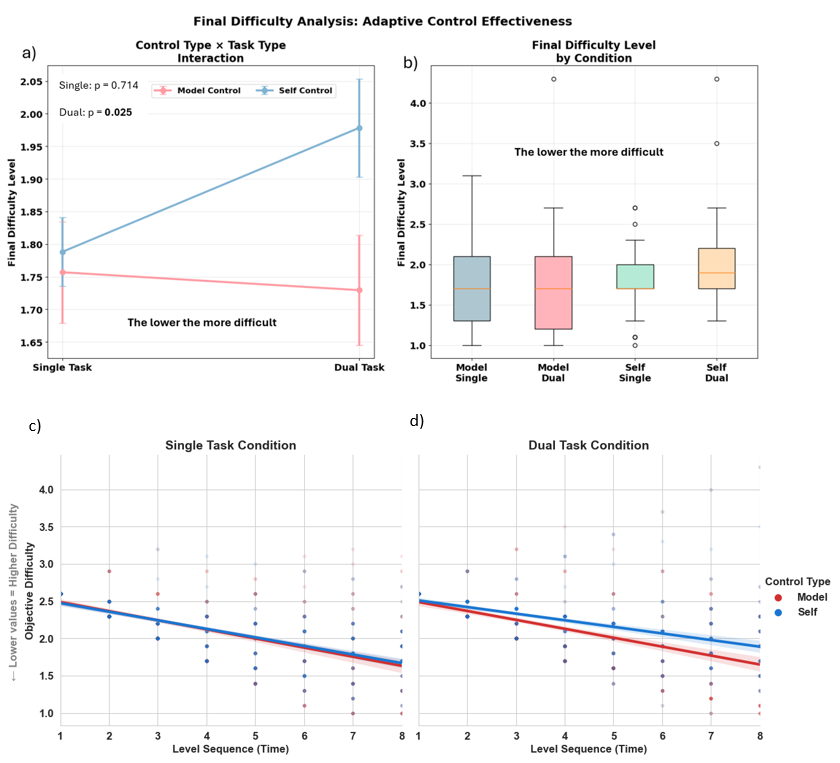}
  \caption{Final difficulty analysis revealing control type effectiveness and task complexity interactions. (a) Control type × task type interaction plot demonstrating equivalent performance in single-task conditions but divergence in dual-task conditions, with model control achieving harder final difficulties. Statistical comparisons show non-significant differences in single-task conditions (p=0.714) but marginally significant differences in dual-task conditions (p=0.025, corrected p=0.05), indicating that model-based control was less effective than self-assessment control when cognitive resources were divided between VR task performance and arithmetic processing; (b) Final difficulty distributions across all four experimental conditions, showing the magnitude of differences between control types within each task complexity level. Lower difficulty values indicate harder task settings. Box plots show median, quartiles, and outliers; error bars in the interaction plot represent standard error of the mean.}
  \label{fig:final_difficulty}
\end{figure}
The individual predictions might not be the most reliable way of evaluating the accuracy of the model, given that they rely on the participant and model agreeing about individual 60s-long level. That method of evaluation, although providing an absolute measure of accuracy, misses out on the general trend in difficulty adjustment. We argue that if the model converges to a similar difficulty level as the participant does by the end of each condition, then it is an indication that the overall cognitive load trend has been captured accurately. Conversely, if the model diverges from the participant-driven difficulty, it should be evidence for failure of the adaptive system. Therefore, we investigated both the temporal dynamics of difficulty adjustment and the final convergence points between conditions.

To capture the continuous nature of this adaptation process, we first analyzed the slope of difficulty changes over time using a Linear Mixed-Effects Model (LMM). The model included fixed effects for Time (Level Sequence), Control Type (Model vs Self), and Task Complexity (Single vs Dual), along with random intercepts and slopes for Time grouped by Participant. Residual diagnostics indicated a minor deviation from normality ($W = 0.953$). This deviation was attributed to the discrete, step-wise nature of the difficulty levels. Given the large number of observations ($N=1,632$) and the balanced study design, the model was deemed robust to these minor violations without the need for transformation. We followed that by a $2 \times 2$ repeated measures ANOVA was conducted on the final level played, with Control Type and Task Complexity as factors. Prior to analysis, the distribution of final difficulty levels was inspected. Shapiro-Wilk tests indicated deviations from normality in all conditions ($W$ range: $0.83 - 0.95$, $p < .05$). Visual inspection of Q-Q plots revealed that these deviations were driven by the discrete nature of the difficulty levels (step-wise increments) rather than extreme outliers. Given the balanced design and sample size ($N=51$), the repeated measures ANOVA was deemed robust to these violations.

The analysis revealed a significant main effect of Time ($\beta = -0.122, SE = 0.009, p < .001$), confirming that objective difficulty generally increased (numerical value decreased) as the session progressed across all conditions. Crucially, a marginally significant three-way interaction was observed between Time, Control Type, and Task Complexity ($\beta = 0.024, SE = 0.013, p = .073$). As visualized in Figure \ref{fig:final_difficulty}c and \ref{fig:final_difficulty}d, the difficulty trajectories diverge under load. While the participants' adaptation slope was aligned with the model's in the single task condition (Figure \ref{fig:final_difficulty}c), it diverged in the dual-task condition (Figure \ref{fig:final_difficulty}d, red line), increasing difficulty at a slower rate than in the model-controlled condition. 

This divergence in adaptation rates accumulated over the session, culminating in a crucial interaction at the final difficulty level. The results of the $2 \times 2$ repeated measures ANOVA revealed no significant main effect of task complexity ($t(50)=1.62$, $p=0.112$), and the main effect of control type approached but did not reach significance ($t(50)=-1.85$, $p=0.070$). Although the omnibus interaction was not statistically significant ($t(50)=-1.79$, $p=0.080$), planned simple main effects analysis confirmed the trend observed in the slope analysis. 

In single-task conditions, model-based and self-assessment control achieved virtually equivalent final difficulty levels (Model: $1.76 \pm 0.55$, Self: $1.79 \pm 0.38$, $t=-0.37$, $p=0.714$), indicating comparable performance when cognitive load was low (see Figure \ref{fig:final_difficulty}a). However, in dual-task conditions, model-based control assigned significantly harder final difficulties than participants selected for themselves (Model: $1.73 \pm 0.60$, Self: $1.98 \pm 0.53$, $t=-2.31$, $p=0.025$, $d=0.33$). After applying a Bonferroni correction for two comparisons, this difference remained significant (corrected $p=0.05$).

The divergence in final difficulty played, along with no difference in weighted game performance across dual tasks, indicates that participants are more conservative with their assessments. They adopted a flatter adaptation slope (Figure \ref{fig:final_difficulty}d) and ended at easier levels than the model determined they were capable of handling but ended up achieving similar weighted game performance and equivalent arithmetic error rates. This suggests that despite the model pushing the participants to play at harder levels in dual task designs, it does not come at a cost in difficulty-weighted performance.

\subsection{Divergence in Cognitive Load Judgments: Human vs. Model Thresholds}

\begin{table*} 
\caption{Cognitive Load Judgment Agreement Analysis for Raw Game Performance and Difficulty-Weighted Game Performance by Task Type and Perceived Difficulty Category - . \textbf{Bold values indicate statistical significance.} *** p < 0.001, ** p < 0.01, * p < 0.05. Positive t-statistics indicate higher self-control game performance relative to model-control.
Cohen's d effect sizes: small (0.2), medium (0.5), large (0.8).}

\centering
\begin{tabular}{@{}llccc|ccc@{}}
\toprule
\multirow{2}{*}{\textbf{Task Type}} & \multirow{2}{*}{\textbf{Category}} & 
\multicolumn{3}{c|}{\textbf{Raw Game Performance}} & 
\multicolumn{3}{c}{\textbf{Weighted Game Performance}} \\
\cmidrule(lr){3-5} \cmidrule(l){6-8}
& & \textit{t-stat} & \textit{p-value} & \textit{Cohen's d} & \textit{t-stat} & \textit{p-value} & \textit{Cohen's d} \\
\midrule
\multirow{3}{*}{Single} 
& Too Easy & 5.189 & \textbf{0.000***} & 0.749 & -1.551 & 0.128 & -0.224 \\
& Just Right & 0.491 & 0.626 & 0.070 & -1.054 & 0.297 & -0.151 \\
& Too Difficult & -5.906 & \textbf{0.000***} & -0.958 & -0.185 & 0.854 & -0.030 \\
\midrule
\multirow{3}{*}{Dual} 
& Too Easy & 4.611 & \textbf{0.000***} & 0.758 & -1.128 & 0.267 & -0.185 \\
& Just Right & 3.241 & \textbf{0.002**} & 0.463 & -3.322 & \textbf{0.002**} & -0.475 \\
& Too Difficult & -2.437 & \textbf{0.019*} & -0.376 & 0.347 & 0.731 & 0.054 \\
\midrule
\multirow{3}{*}{Pooled} 
& Too Easy & 5.146 & \textbf{0.000***} & 0.743 & -0.998 & 0.324 & -0.144 \\
& Just Right & 2.136 & \textbf{0.038*} & 0.299 & -3.972 & \textbf{0.000***} & -0.556 \\
& Too Difficult & -5.080 & \textbf{0.000***} & -0.749 & 1.185 & 0.242 & 0.175 \\
\bottomrule

\end{tabular}
\label{tab:performance_analysis}
\end{table*}
\begin{figure} 
    \centering
    \includegraphics[width=1\textwidth]{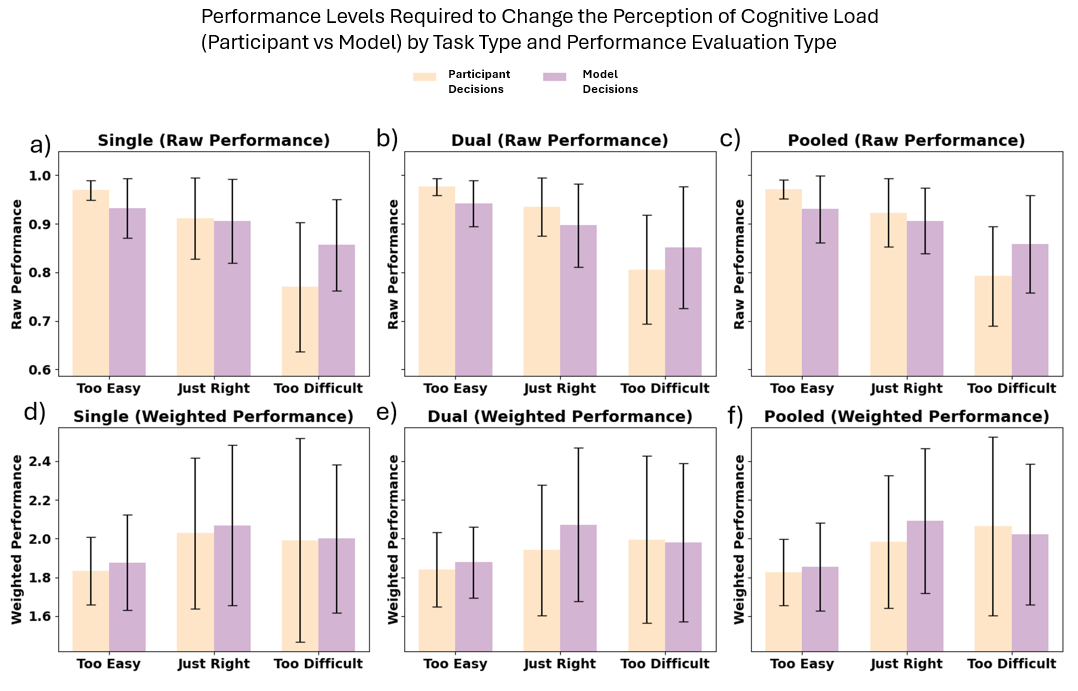}
    \caption{Game performance analysis showing the relationship between difficulty judgments/predictions and actual game performance. The 2×3 grid compares raw performance scores (top row) and weighted game performance scores (bottom row) for both model and self-driven conditions, including an additional breakdown for single (left column) and dual tasks (middle column) as well as both conditions pooled (right column). Boxplots display the distribution of mean performance scores aggregated by participant and difficulty (\textit{Too Easy, Just Right and Too Difficult}). This visualization presents the mapping of raw and weighted game performance across difficulty estimations.}
    \label{fig:performance_by_judgment}
\end{figure}

\begin{figure} 
    \centering
    \includegraphics[width=1\textwidth]{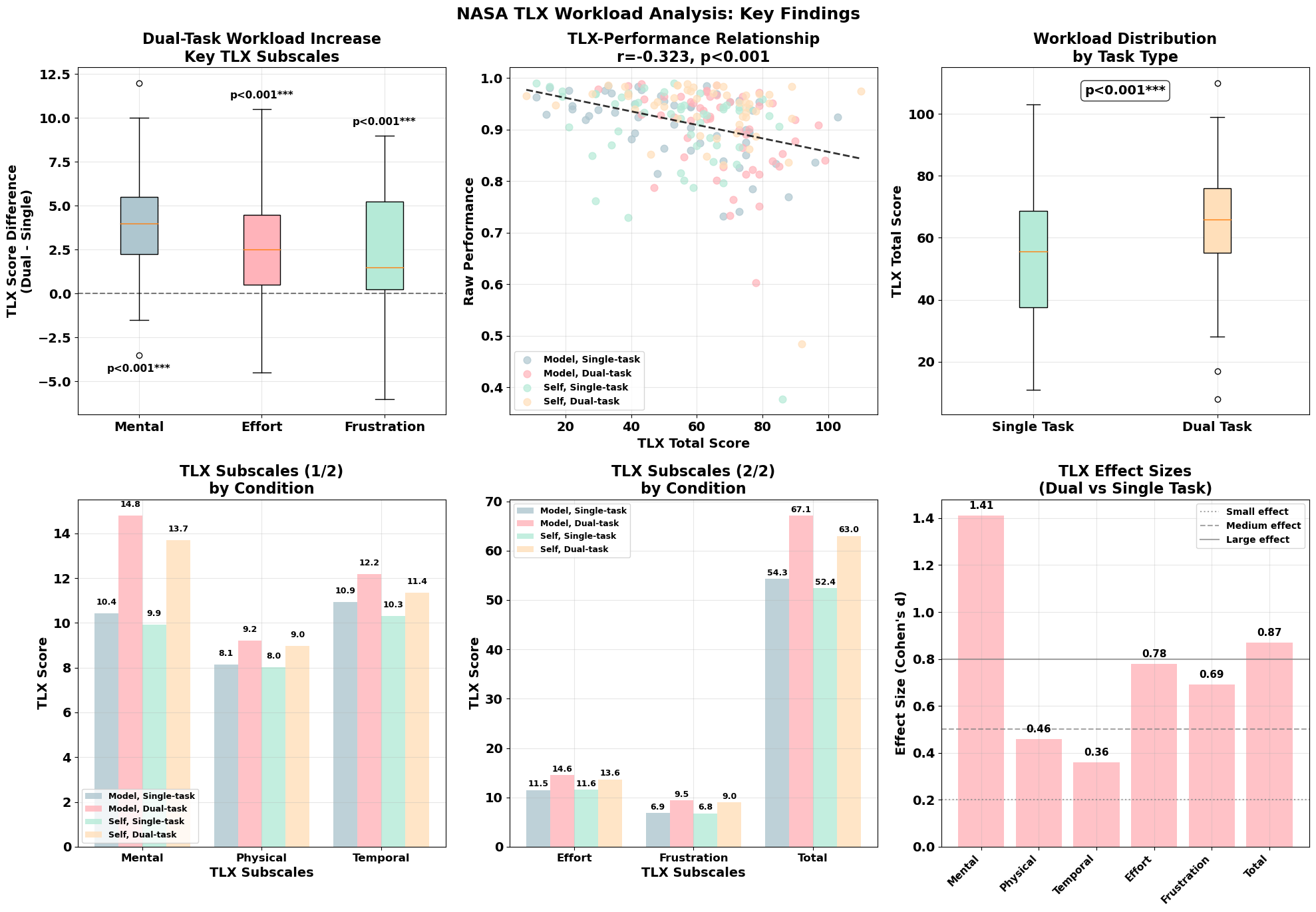}
    \caption{NASA TLX workload analysis revealing dual-task cognitive burden and performance relationships. (a) Dual-task workload increases showing significant elevations in mental demand ($p<0.001$, $d=1.41$), effort ($p<0.001$, $d=0.78$), and frustration ($p<0.001$, $d=0.69$) when comparing dual-task to single-task conditions within participants; (b) TLX-performance correlation demonstrating a significant negative relationship ($r=-0.323$, $p<0.001$) between subjective workload and spawn-destroy task performance across all conditions; (c) Total workload distribution confirming significantly higher TLX scores in dual-task compared to single-task conditions ($p<0.001$, $d=0.87$); (d) TLX subscale profiles for mental demand, physical demand, and temporal demand across the four experimental conditions, with Model Dual-task showing the highest mental demand ratings; (e) TLX subscale profiles for effort, frustration, and total workload, demonstrating consistent patterns where dual-task conditions (Model Dual-task and Self Dual-task) exceed single-task conditions across all measures; (f) Effect size analysis quantifying the magnitude of dual-task workload increases, with mental demand showing the largest effect ($d=1.41$), followed by total workload ($d=0.87$), effort ($d=0.78$), and frustration ($d=0.69$), while physical and temporal demands showed smaller but significant effects ($d=0.46$ and $d=0.36$, respectively). Box plots display median, quartiles, and outliers; bar charts show condition means with individual values labeled; effect size reference lines indicate Cohen's conventions for small (0.2), medium (0.5), and large (0.8) effects.}
    \label{fig:tlx_workload}
\end{figure}

Analysis of the cognitive load judgment thresholds revealed systematic differences in how the computational model and human participants categorised task difficulty, providing insight into the reasons behind prediction discrepancies observed throughout the experiment (Figure~\ref{fig:performance_by_judgment}). Specifically, we examined the mean game performance levels associated with each difficulty label (\textit{Too Easy, Just Right and Too Difficult}) to compare how model predictions relate to participant perceptions of optimum difficult levels, see Table \ref{tab:performance_analysis}. In the raw game performance analysis, participants tended to overestimate their ability when they perceived a task as \textit{Too Easy} and underestimate it when they judged it as \textit{Too Difficult}. These systematic biases disappeared in the weighted game performance analysis, where game performance was adjusted for objective task difficulty, suggesting that participants did not fully factor in difficulty when forming their judgments. By contrast, the computational model maintained consistent judgment thresholds across conditions, with differences persisting only in the \textit{Just Right} category, which is consistent with the training strategy prioritising \textit{Too Easy} and \textit{Too Difficult} classification over \textit{Just Right}. This highlights the model’s greater objectivity in accounting for task demands, while participant judgments remained more subjective and biased by a shallower sense of success perception determined by the amount of cubes left at the end of the level regardless of how many of them appeared.

\subsection{Subjective Workload Assessment with NASA TLX}

While real-time cognitive workload evaluation based on categorical difficulty labels (\textit{Too Easy, Just Right, Too Difficult}) provides insight into moment-to-moment task adaptation, it may not fully capture the broader experiential and affective dimensions of user engagement. To complement this model-driven assessment, we employed the NASA Task Load Index (TLX) at the end of each experimental condition.

To examine the effects of the adaptive system and the secondary task on subjective experience, a $2 \times 2$ repeated measures ANOVA was conducted on the weighted TLX scores, with Control Type (Model vs. Self) and Task Complexity (Single vs. Dual) as within-subject factors. Shapiro-Wilk tests confirmed that NASA-TLX workload scores satisfied the assumption of normality across all four experimental conditions ($W > 0.96$, $p > .19$), validating the use of parametric analysis without correction.

\subsubsection{Task Complexity}
The analysis revealed a robust main effect of Task Complexity across nearly all workload dimensions, confirming the efficacy of the arithmetic task in inducing cognitive load. For the overall Total Workload score, participants reported significantly higher burden in the dual-task condition compared to the single-task condition ($F(1, 50) = 38.48, p < .001$). 

Breakdown by subscale indicated that this increase was driven primarily by \textbf{Mental Demand} ($F(1, 50) = 101.03, p < .001$) and \textbf{Effort} ($F(1, 50) = 31.30, p < .001$), reflecting the high cognitive cost of concurrent arithmetic processing. \textbf{Frustration} levels were also significantly higher in the dual-task condition ($F(1, 50) = 24.28, p < .001$), suggesting that the divided attention requirement negatively impacted affective state. \textbf{Physical Demand} ($F(1, 50) = 10.87, p = .002$) and \textbf{Temporal Demand} ($F(1, 50) = 6.45, p = .014$) showed smaller but statistically significant increases. notably, the \textbf{Performance} subscale did not differ significantly between task conditions ($F(1, 50) = 2.40, p = .127$), indicating that participants felt their level of success was comparable regardless of the added complexity.

\subsubsection{Control Type and Interaction Effects}
In contrast to the strong effects of task complexity, the main effect of Control Type was not statistically significant for Total Workload ($F(1, 50) = 1.87, p = .177$), nor for any individual subscale (all $p > .08$). This indicates that participants did not perceive the Model-controlled adaptive difficulty as significantly more demanding or intrusive than their own self-regulated adjustments.

Crucially, the interaction between Control Type and Task Complexity was not significant for Total Workload ($F(1, 50) = 0.27, p = .608$) or any subscale. This suggests that the increase in workload observed during the dual-task condition was consistent across both control methods. The adaptive model did not exacerbate the workload of the secondary task, nor did it mitigate it significantly more than users could achieve on their own.

\subsubsection{Game Performance Relationships with NASA TLX-based Workload Ratings}

Significant negative correlations emerged between TLX ratings and task performance, particularly for both raw and weighted game performance measures. For raw game performance, strong negative correlations were observed with performance perception ($r=-0.442$, $p<0.001$), frustration ($r=-0.399$, $p<0.001$), and total workload ($r=-0.323$, $p<0.001$). For weighted game performance, which accounts for task difficulty, the correlations were weaker but still significant for game performance perception ($r=-0.287$, $p=0.002$), frustration ($r=-0.245$, $p=0.008$), and total workload ($r=-0.206$, $p=0.023$). Other TLX subscales showed non-significant correlations with weighted game performance, including mental demand ($r=-0.152$, $p=0.103$), physical demand ($r=-0.087$, $p=0.346$), and temporal demand ($r=-0.114$, $p=0.208$). These relationships suggest that higher subjective workload was associated with poorer game performance, with raw game performance showing stronger associations than weighted game performance. Additionally, mental demand showed a notable positive correlation with arithmetic error rates ($r=0.353$, $p<0.001$), confirming that dual-task interference was reflected in both objective game performance decrements and subjective cognitive load ratings.
 
\subsection{Learning and Fatigue Effects}

To assess participant adjustment to the task over time, we analysed learning and fatigue patterns by comparing participants' game performance and NASA-TLX results between their first and second exposure to each task type (single-task vs dual-task conditions). The experiment design ensured consistent sequencing across participants. Single task blocks were always in positions 1 and 3, while dual task blocks appeared in positions 2 and 4. This fixed ordering allowed for direct within-subject comparisons, controlling for time-on-task and interval effects across all participants. Repeated-measures ANOVA was conducted separately for each task type with single-task results presented in Table \ref{tab:single_learning} and dual-task results presented in Table \ref{tab:dual_learning}. 
\begin{table} 

\centering
\caption{Repeated-Measures ANOVA Results for Single-Task Conditions}
\begin{tabular}{lcccc}
\toprule
Metric & F Value & Num DF & Den DF & p-value \\
\midrule
Raw Game Performance & 3.3118 & 1 & 50 & 0.0748 \\
Weighted Game Performance & 38.2463 & 1 & 50 & <0.0001 \\
Final Difficulty & 14.3117 & 1 & 50 & 0.0004 \\
TLX Total & 3.8970 & 1 & 50 & 0.0539 \\
TLX Mental & 5.3976 & 1 & 50 & 0.0243 \\
TLX Physical & 0.1066 & 1 & 50 & 0.7454 \\
TLX Temporal & 3.8713 & 1 & 50 & 0.0547 \\
TLX Perceived Performance & 4.0689 & 1 & 50 & 0.0491 \\
TLX Effort & 8.0692 & 1 & 50 & 0.0065 \\
TLX Frustration & 0.3982 & 1 & 50 & 0.5309 \\
\bottomrule
\label{tab:single_learning}
\end{tabular}
\end{table}

\begin{table} 
\centering
\caption{Repeated-Measures ANOVA Results for Dual-Task Conditions}
\begin{tabular}{lcccc}
\toprule
Metric & F Value & Num DF & Den DF & p-value \\
\midrule
Raw Game Performance & 6.1670 & 1 & 50 & 0.0164 \\
Weighted Game Performance & 14.4091 & 1 & 50 & 0.0004 \\
Final Difficulty & 2.6869 & 1 & 50 & 0.1075 \\
TLX Total & 12.7224 & 1 & 50 & 0.0008 \\
TLX Mental & 20.0635 & 1 & 50 & <0.0001 \\
TLX Physical & 5.3026 & 1 & 50 & 0.0255 \\
TLX Temporal & 3.6519 & 1 & 50 & 0.0617 \\
TLX Perceived Performance & 7.9402 & 1 & 50 & 0.0069 \\
TLX Effort & 9.5341 & 1 & 50 & 0.0033 \\
TLX Frustration & 3.2905 & 1 & 50 & 0.0757 \\
\bottomrule
\label{tab:dual_learning}

\end{tabular}
\end{table}

\subsubsection{Performance Dynamics and Skill Acquisition}

To assess the sustainability of performance and the presence of learning effects over the course of the session, we analysed the trajectory of difficulty-weighted performance. A Linear Mixed-Effects Model (LMM) was fitted with \textbf{Level Sequence} (Time), \textbf{Control Type}, and \textbf{Task Complexity} as fixed effects, including random intercepts and slopes for participants. Prior to analysis, residual diagnostics were performed to validate the model. The Shapiro-Wilk test on model residuals indicated an excellent fit to the normal distribution ($W = 0.997$), and visual inspection of Q-Q plots confirmed that residuals closely followed the theoretical diagonal. A scatterplot of residuals versus fitted values displayed a random dispersion of points, satisfying the homoscedasticity assumption required for linear mixed-effects modeling.

\begin{figure}
  \centering
  \includegraphics[width=1\textwidth]{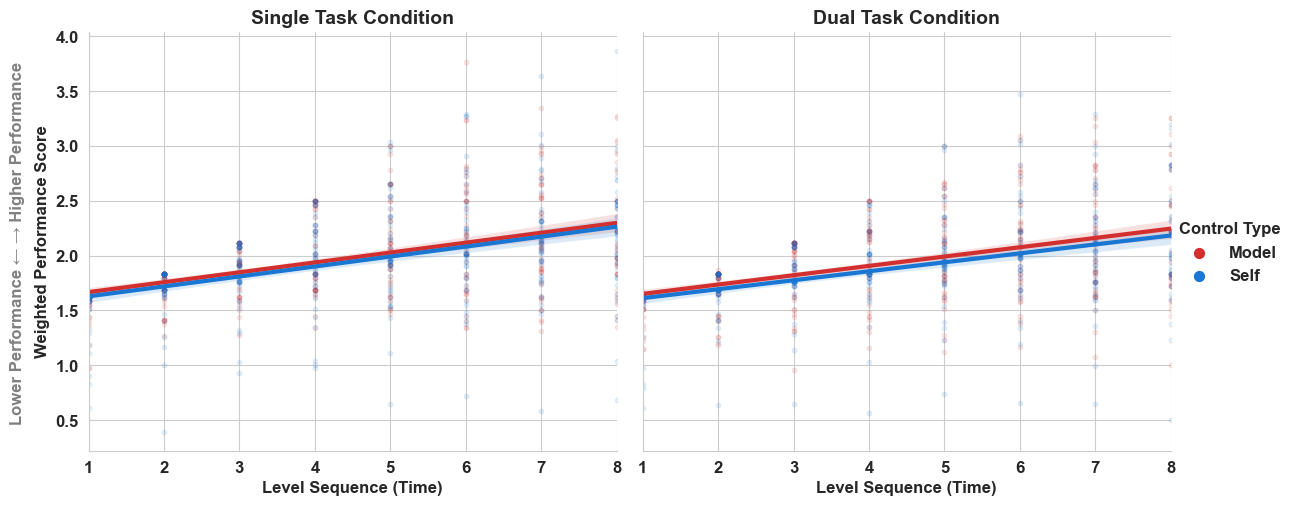}
  \caption{Linear regression of Weighted Performance trajectories over time. (a) Single Task condition (left) and (b) Dual Task condition (right). Red lines represent Model Control; Blue lines represent Self Control. The significant positive slope across all conditions ($p < .001$) indicates a consistent learning effect. Shaded regions represent 95\% confidence intervals.}
  \label{fig:learning_curves}
\end{figure}

The analysis revealed a robust and statistically significant main effect of \textbf{Time} (Level Sequence) on performance ($\beta = 0.090, SE = 0.009, p < .001$). As illustrated in Figure \ref{fig:learning_curves}, participants in all conditions demonstrated a steady improvement in weighted performance as the session progressed, indicating a continuous learning effect rather than fatigue. The positive slope suggests that users were able to handle increasingly difficult challenges (or maintain accuracy at higher difficulty levels) without a degradation in capacity.

This learning rate was consistent across all experimental conditions. The interaction between Time and Control Type was not significant ($p = .942$), nor was the three-way interaction between Time, Control Type, and Task Complexity ($p = .722$). This indicates that the rate of improvement was identical regardless of whether the difficulty was controlled by the model or the user, and regardless of the secondary task load.

While the model-controlled condition did not induce a \textit{steeper} learning curve, it successfully supported the same rate of skill acquisition as in the self-controlled conditions. The absence of a negative interaction effect in the dual-task condition further confirms that the additional cognitive load did not inhibit the users' ability to learn and improve at the primary task over time.

This continuous learning trajectory culminated in significant performance retention between the first and second exposures of each task. Repeated-measures ANOVA confirmed that Weighted Game Performance improved significantly in the second exposure for both Single-Task ($\Delta = +0.224, p < .001, d = 0.87$) and Dual-Task ($\Delta = +0.146, p < .001, d = 0.53$) conditions. 

Raw Game Performance (unweighted) showed a more modest improvement, reaching significance only in the Dual-Task condition ($p = .016$). This discrepancy between weighted and raw metrics highlights the function of the adaptive system: as users learned (positive slope), the system increased the difficulty to match their new capacity. Consequently, while their raw score remained relatively stable, their difficulty-weighted performance increased because they were operating at a significantly higher difficulty level. This is confirmed by the analysis of Final Difficulty, which significantly increased (numerical value decreased) in the second Single-Task exposure ($p < .001, d = 0.53$), indicating that the learning observed in the LMM was successfully converted into higher challenge acceptance.

\subsubsection{Workload Reduction Over Time}

\begin{figure} 
    \centering
    \includegraphics[width=1\textwidth]{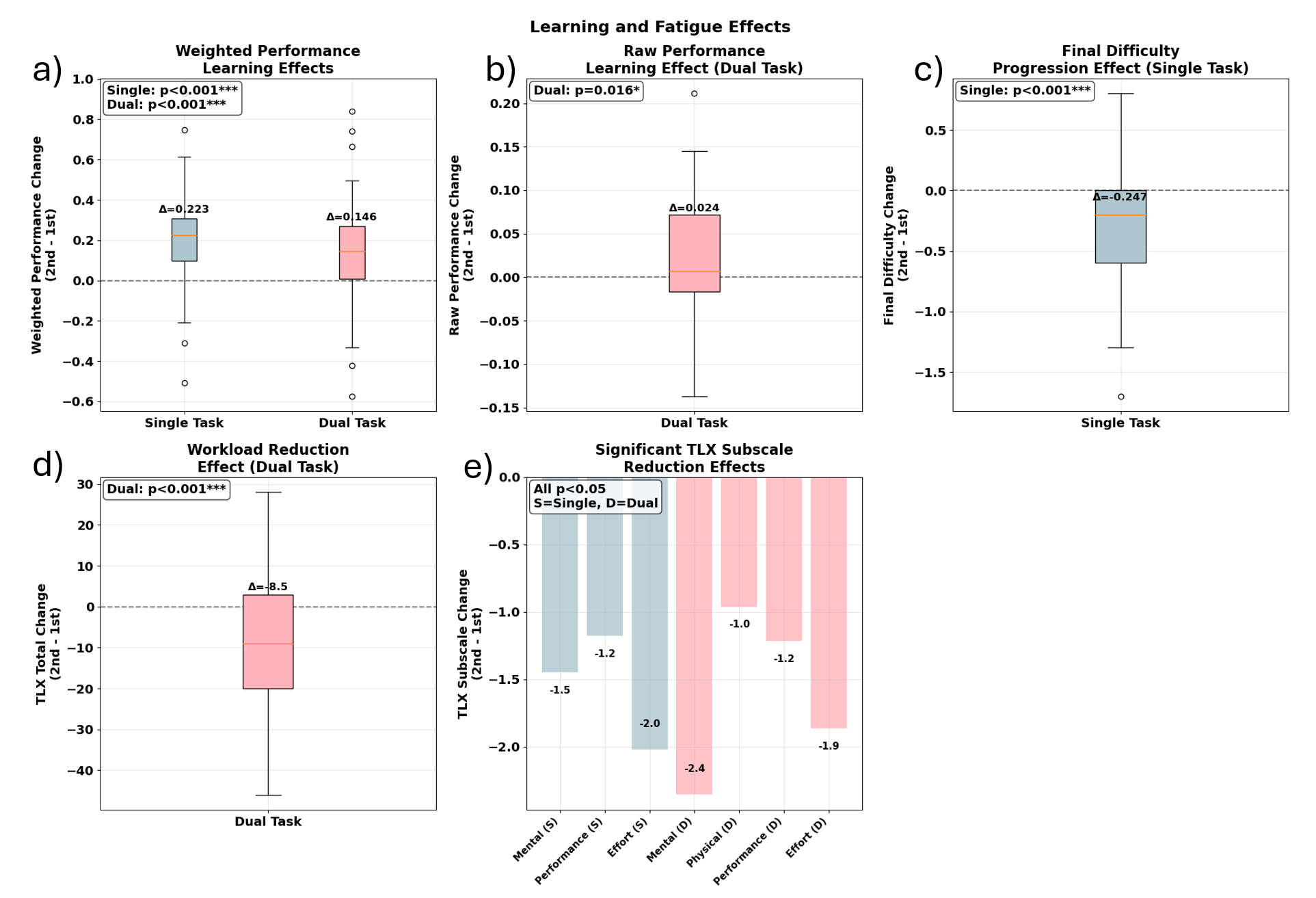}
    \caption{Significant learning and fatigue effects across experimental conditions. The figure displays five key findings: (a) Weighted game performance improvements in both single-task ($\Delta$=0.224, p<0.001) and dual-task conditions ($\Delta$=0.146, p<0.001); (b) Raw game performance improvement in dual-task conditions only ($\Delta$=0.024, p=0.016); (c) Final difficulty level reduction in single-task conditions only ($\Delta$=-0.162, p<0.001), indicating progression to harder tasks; (d) Overall workload reduction in dual-task conditions ($\Delta$=-8.5, p<0.001); and (e) Specific TLX subscale reductions, with single-task conditions showing improvements in mental demand ($\Delta$=-1.5, p=0.024), performance perception ($\Delta$=-1.2, p=0.049), and effort ($\Delta$=-2.0, p=0.007), while dual-task conditions demonstrated broader improvements across mental demand ($\Delta$=-2.4, p<0.001), physical demand ($\Delta$=-1.0, p=0.026), performance perception ($\Delta$=-1.2, p=0.007), and effort ($\Delta$=-1.9, p=0.003) subscales (all p<0.05). Box plots show median, quartiles, and outliers; bar charts display mean differences with statistical significance levels indicated.}    
    \label{fig:learning_fatigue}
\end{figure}

The most pronounced learning effect was observed in subjective workload measures. Total TLX scores decreased significantly for dual-task conditions by 8.5 points ($p < 0.001, d = 0.499$), as illustrated in the individual trajectory plot (Figure~\ref{fig:learning_fatigue}d). This substantial workload reduction was driven by improvements across multiple TLX subscales (Figure~\ref{fig:learning_fatigue}e), with mental demand showing the largest decrease (-2.4 points, p < 0.001), followed by effort (-1.9 points, p = 0.003), perceived performance (-1.2 points, p = 0.007), and physical demand (-1.0 points, p = 0.026).
Single-task conditions showed more modest workload reductions, with significant decreases in mental demand (-1.5 points, p = 0.024), effort (-2.0 points, p = 0.007), and perceived performance (-1.2 points, p = 0.049), while total TLX scores approached significance (p = 0.054).

\subsection{Individual Differences in Real-time Model Performance}

Analysis of per-participant model accuracy revealed substantial individual variability in prediction success, ranging from 6.2\% to 59.4\% accuracy across the 51 participants (M = 36.7\%, SD = 12.0\%). This wide distribution suggests that while the model performs modestly on average, it achieves high accuracy for specific participant subgroups while failing entirely for others.

\subsubsection{Fairness Analysis of Demographic Predictors in Model Performance}

 Male participants demonstrated superior model accuracy compared to female participants (40.3\% vs 34.8\%), though this difference did not reach statistical significance (t = 1.565, p = 0.124). Age showed no relationship with model accuracy (r = 0.022, p = 0.877), with equivalent performance across age groups (18-25: 35.5\%, 26-35: 37.5\%, 36+: 37.8\%, F = 0.188, p = 0.829). Handedness effects could not be reliably assessed due to the small number of left-handed participants (n = 2).

\subsubsection{Predictive Factors for Model Success}

Correlation analysis identified probability-accuracy calibration as the strongest individual predictor of overall model performance (r=0.385,p=0.005). In the context of the zero-calibration deployment strategy, this metric served as a representation of the distributional alignment between an individual participant and the global training norms. Participants for whom the model performed best demonstrated positive correlations between model confidence and accuracy, indicating that their standardized physiological features aligned well with the population-level distributions used for scaling. Conversely, low-performing participants exhibited weak or negative correlations, suggesting a distributional shift where the global scaler failed to normalize their unique physiological baselines effectively.

In contrast to this physiological alignment metric, demographic variables (age, gender), baseline VR performance, and subjective workload measures showed only weak non-significant associations with model accuracy. A predictive model incorporating these readily observable characteristics explained only 13.7\% of the variance in individual model accuracy. This indicates that traditional screening criteria provide limited predictive power for identifying suitable candidates for model-based adaptive control, as performance is primarily determined by the compatibility between the user's physiology and the global normalization model.

\section{Discussion}

\subsection{Methodological Contributions}

A prevalent methodological shortcut in the field of cognitive load estimation is the use of 'task stacking' (adding discrete secondary tasks) as a proxy for cognitive load manipulation (\cite{clare2025, omnicept}). While this creates distinct physiological classes, it prevents the granular, incremental adjustments required for effective training within a single task. Moreover, in particular when using eye-tracking as means of detecting cognitive load shifts, task stacking can result in the model learning to distinguish the eye-tracking patterns required to perform the additional tasks, rather than the actual cognitive load increase. In the domain of adaptive design, existing systems often rely on simple game-based heuristics (\cite{araujo2019, drey2020}), modulate visual complexity rather than intrinsic load (\cite{lindlbauer2019context, chiossi2023adapting}), or focus on affective states rather than cognitive capacity (\cite{pinilla2023real, quintero2025personalized}).

Furthermore, real-time deployment of cognitive load models requires more than high offline accuracy. Current research prioritises methodological choices that maximize offline performance but they do not consider real-world applicability. For instance, the majority of studies estimating cognitive load from eye-tracking and physiological data rely on within-participant cross-validation, which inflates accuracy by training and testing on the same individuals (\cite{shojaeizadeh2019detecting, skaramagkas2021cognitive, he2022classification, omnicept}). When studies adopt leave-one-participant-out validation, which is a more realistic test of generalization, performance drops significantly (\cite{bachurina2022multiple, clare2025}).

The current study addresses these gaps by presenting the first empirical validation of a real-time, closed-loop adaptive VR system driven exclusively by eye-tracking and physiological data. We utilized a rigorous leave-one-out cross-validation protocol for the deployment model, manipulated difficulty intrinsically within the task to allow for smooth incremental adjustments, and deployed the system using consumer-grade hardware suitable for at-home use. As noted in a recent systematic review by \cite{mortazavi2024dynamic}, the field is shifting from rule-based heuristics to data-driven methods capable of capturing internal states; our work represents a practical realisation of this shift, moving from offline analysis to live, closed-loop adaptation.

\subsection{Key Findings}

\subsubsection{Conservatism Bias vs. Overconfidence: The Divergence of Objective and Subjective Difficulty}

A critical finding of this study is the divergence in difficulty regulation strategies under high cognitive load. In dual-task conditions, the model drove participants to significantly harder difficulty levels than they selected for themselves. Crucially, this increase in difficulty did not result in a performance penalty (difficulty-weighted game performance remained equivalent as did arithmetic error rates). We consider this to indicate a conservatism bias, where participants in the dual task condition underestimate their remaining capacity and self-regulate toward easier, safer difficulty levels.

This finding contrasts with \cite{constant2019dynamic}, who observed that DDA systems typically result in player \textit{overconfidence}. However, their study employed a betting mechanism that explicitly incentivised risk assessment in a single-task context. The players needed to adjust the bet amount prior to each round, knowing that if they win, they will get the bet back along with a reward multiplier and if they lose, the money they bet will be taken away. As noted by \cite{guo2025exploratory}, such betting designs rely on \textit{prediction} (future expectation), which typically triggers optimism biases, whereas post-hoc ratings or physiological monitoring rely on \textit{reflection} (past evaluation), which often yields more rational or conservative estimates.
In our study, the hidden nature of the control loop removed this gamification incentive as players were unaware that their difficulty judgments would influence the task difficulty. Instead, the observed conservatism is better explained by the dissociation theory (\cite{yeh1988dissociation}). Yeh and Wickens posit that subjective ratings reflect the cognitive resources invested to maintain performance, rather than the performance outcome itself. Consequently, participants likely felt the increased metabolic cost of the dual-task and rated it as \textit{Too Difficult}, even though their objective performance output remained stable. Furthermore, this judgment was likely reinforced by a metacognitive rule (\cite{hendy1993measuring}), where participants assumed that the structural presence of a secondary task necessitated higher difficulty ratings regardless of their actual remaining capacity.

This heuristic reliance confirms the fundamental disconnect between Objective Game Difficulty and Subjective Game Difficulty proposed by \cite{guo2025exploratory} and \cite{constant2017objective}. Our data indicates that participants anchored their judgments on raw performance (absolute number of cubes they missed despite difficulty fluctuations), whereas the model aligned with difficulty-weighted performance. This further supports the metacognitive rule proposed by \cite{hendy1993measuring}, where participants focus on observable cues, in this case using raw performance as a mental shortcut to judge the difficulty of the task. By overriding the user's subjective model ('two tasks must be harder than one') with physiological reality, the system successfully prioritised the training goal over user comfort in dual task conditions. This provides a compelling methodological argument against using task stacking as a proxy for cognitive load, as it introduces structural task biases that obstruct participants' objectivity.

\subsubsection{Adaptive Smoothing: The Shift to \textit{Just Right}}
We observed a significant distributional shift in difficulty ratings between the offline data collection (Experiment 1) and the real-time deployment (Experiment 2). In the real-time study, participant ratings converged heavily on \textit{Just Right}, with a significant reduction in \textit{Too Easy} and \textit{Too Difficult} edge cases. The adaptive system stabilises the difficulty through informed and incremental difficulty adjustments as opposed to random and large amplitudes observed in fixed difficulty systems (\cite{mortazavi2024dynamic}). Even though the classification model itself was pre-trained on a fixed diffuculty task, the closed-loop nature of the system naturally minimised the occurrence of non-optimal states.

This aligns with the goal-based DDA framework proposed by \cite{guo2024rethinking}, which suggests that the primary value of DDA is its ability to control the difficulty progression precisely to achieve a 'design curve', which in our case aims to align the \textit{Just Right} states with peak difficulty-weighted performance. By continuously correcting deviations, our system smoothed the user's experience, effectively maintaining them within the Flow Channel \cite{guo2024rethinking} far more consistently than the random or fixed-step progressions used in the offline experiment. This confirms that even a moderately accurate classifier, when embedded in a responsive loop, can stabilize the user experience and mitigate the edge states of boredom and frustration.

\subsubsection{Model Utility: Automating the Training Loop}
Finally, our results demonstrate the utility of physiological adaptation as a surrogate for manual control. The model-driven conditions matched the participant-driven conditions across all key metrics: weighted game performance, subjective workload (NASA-TLX), and learning rates. This equivalence suggests that the explicit step of asking users for difficulty assessments, which breaks immersion and is subject to the aforementioned biases, can be effectively replaced by model-driven monitoring of their physiological states.

This validation is crucial for the advancement of serious games and cognitive training. As \cite{guo2024rethinking} argue, serious games often have characterizing goals (such as maximizing training capacity) that differ from pure entertainment goals. Our model successfully prioritised this serious goal in the dual-task condition by pushing users beyond their comfort zone, without causing the frustration or disengagement that \cite{mortazavi2024dynamic} warn against. By demonstrating that an automated system can maintain the same learning trajectory and workload profile as a user-controlled one, we provide a blueprint for autonomous training systems that are robust to user fatigue or poor self-assessment.

\subsection{Limitations and Future Directions}
We observed substantial individual variability in model performance, consistent with the individual differences often cited in physiological computing. While the model achieved high accuracy for a subset of users, it performed closer to chance for others. This suggests that the disconnect between physiology and perception is not uniform; some users may have a stronger alignment between their internal state and their subjective reporting than others. Future iterations of this system would benefit from a calibration phase to align the model's decision boundaries with the individual's baseline physiology, rather than relying on global normalization. Alternatively, a larger and more diverse offline dataset could result in a more universal model.

\subsection{Conclusions}
We argue that the transition from offline model training to real-time deployment is not straightforward. It requires eradicating methodological shortcuts such as within-participant cross-validation and task stacking that inflate offline accuracy but hinder real-world applicability. This study addresses these gaps by presenting the first empirical validation of a real-time, closed-loop VR system driven by eye-tracking and physiological metrics.

We demonstrate that an automated system can modulate difficulty as effectively as user self-control, achieving equivalent learning rates and subjective workload. More importantly, in complex dual-task scenarios, our model overcame the participants' tendency toward 'conservatism bias', pushing them to higher difficulty levels without performance penalties. This confirms that physiological loops can sustain a user’s optimal challenge zone more effectively than the users themselves, particularly when cognitive resources are split.

A critical implication of our findings is the fundamental disconnect between subjective perception and objective physiology. As supported by recent theoretical frameworks (\cite{guo2025exploratory, constant2017objective}), players often misjudge their own capacity, anchoring on raw error rates rather than difficulty-weighted performance. This reveals a significant risk in training physiological models purely on subjective labels: users are often unreliable judges of their own cognitive load, in particular in adaptive designs. Consequently, future designers of adaptive training systems must make a strategic choice: optimise for \textit{perceived comfort} (yielding to user bias) or \textit{objective performance} (challenging that bias).

\printcredits

\bibliographystyle{cas-model2-names}

\bibliography{cas-refs}

@article{RN10,
   author = {Bachurina, Valentina and Sushchinskaya, Svetlana and Sharaev, Maxim and Burnaev, Evgeny and Arsalidou, Marie},
   title = {A machine learning investigation of factors that contribute to predicting cognitive performance: Difficulty level, reaction time and eye-movements},
   journal = {Decision Support Systems},
   volume = {155},
   pages = {113713},
   ISSN = {0167-9236},
   DOI = {https://doi.org/10.1016/j.dss.2021.113713},
   url = {https://www.sciencedirect.com/science/article/pii/S0167923621002232},
   year = {2022},
   type = {Journal Article}
}

@inbook{RN11,
   author = {Cohen, Ronald A.},
   title = {Continuous Performance Tests},
   booktitle = {Encyclopedia of Clinical Neuropsychology},
   editor = {Kreutzer, Jeffrey S. and DeLuca, John and Caplan, Bruce},
   publisher = {Springer New York},
   address = {New York, NY},
   pages = {699-701},
   ISBN = {978-0-387-79948-3},
   DOI = {10.1007/978-0-387-79948-3_1280},
   url = {https://doi.org/10.1007/978-0-387-79948-3_1280},
   year = {2011},
   type = {Book Section}
}

@article{he2022classification,
  title={Classification of driver cognitive load: Exploring the benefits of fusing eye-tracking and physiological measures},
  author={He, Dengbo and Wang, Ziquan and Khalil, Elias B and Donmez, Birsen and Qiao, Guangkai and Kumar, Shekhar},
  journal={Transportation research record},
  volume={2676},
  number={10},
  pages={670--681},
  year={2022},
  publisher={SAGE Publications Sage CA: Los Angeles, CA}
}

@article{lundberg2017unified,
  title={A unified approach to interpreting model predictions},
  author={Lundberg, Scott M and Lee, Su-In},
  journal={Advances in neural information processing systems},
  volume={30},
  year={2017}
}

@article{shojaeizadeh2019detecting,
  title={Detecting task demand via an eye tracking machine learning system},
  author={Shojaeizadeh, Mina and Djamasbi, Soussan and Paffenroth, Randy C and Trapp, Andrew C},
  journal={Decision Support Systems},
  volume={116},
  pages={91--101},
  year={2019},
  publisher={Elsevier}
}

@inproceedings{skaramagkas2021cognitive,
  title={Cognitive workload level estimation based on eye tracking: A machine learning approach},
  author={Skaramagkas, Vasileios and Ktistakis, Emmanouil and Manousos, Dimitris and Tachos, Nikolaos S and Kazantzaki, Eleni and Tripoliti, Evanthia E and Fotiadis, Dimitrios I and Tsiknakis, Manolis},
  booktitle={ IEEE International Conference on Bioinformatics and Bioengineering},
  pages={1--5},
  year={2021}
}

@misc{LSL,
  author = {Stenner, Tristan},
  title = {Lab Streaming Layer (LSL) - A software framework for synchronizing a large array of data collection and stimulation devices.},
  year = {2022},
  url = {https://github.com/sccn/labstreaminglayer}
}

@misc{PyS,
  author = {Magel, Lukassmithjer},
  title = {PyShimmer},
  year = {2023},
  url = {https://github.com/seemoo-lab/pyshimmer}
}

@article{cochrane2020load,
  title={Load effects in attention: Comparing tasks and age groups},
  author={Cochrane, Aaron and Simmering, Vanessa and Green, C Shawn},
  journal={Attention, Perception, \& Psychophysics},
  volume={82},
  pages={3072--3084},
  year={2020},
  publisher={Springer}
}

@article{rao2020predicting,
  title={Predicting cognitive load and operational performance in a simulated marksmanship task},
  author={Rao, Hrishikesh M and Smalt, Christopher J and Rodriguez, Aaron and Wright, Hannah M and Mehta, Daryush D and Brattain, Laura J and Edwards III, Harvey M and Lammert, Adam and Heaton, Kristin J and Quatieri, Thomas F},
  journal={Frontiers in Human Neuroscience},
  volume={14},
  pages={222},
  year={2020},
  publisher={Frontiers Media SA}
}

@article{hamilton2023sustained,
  title={Sustained attention in mild cognitive impairment with Lewy bodies and Alzheimer’s disease},
  author={Hamilton, Calum A and Gallagher, Peter and Ciafone, Joanna and Barnett, Nicola and Barker, Sally AH and Donaghy, Paul C and O’Brien, John T and Taylor, John-Paul and Thomas, Alan J},
  journal={Journal of the International Neuropsychological Society},
  pages={1--7},
  year={2023},
  publisher={Cambridge University Press}
}

@article{liu2017multisubject,
  title={Multisubject “learning” for mental workload classification using concurrent EEG, fNIRS, and physiological measures},
  author={Liu, Yichuan and Ayaz, Hasan and Shewokis, Patricia A},
  journal={Frontiers in human neuroscience},
  volume={11},
  pages={389},
  year={2017},
  publisher={Frontiers Media SA}
}

@article{aygun2022investigating,
  title={Investigating methods for cognitive workload estimation for assistive robots},
  author={Aygun, Ayca and Nguyen, Thuan and Haga, Zachary and Aeron, Shuchin and Scheutz, Matthias},
  journal={Sensors},
  volume={22},
  number={18},
  pages={6834},
  year={2022},
  publisher={MDPI}
}

@article{bitkina2021ability,
  title={The ability of eye-tracking metrics to classify and predict the perceived driving workload},
  author={Bitkina, Olga Vl and Park, Jaehyun and Kim, Hyun K},
  journal={International Journal of Industrial Ergonomics},
  volume={86},
  pages={103193},
  year={2021},
  publisher={Elsevier}
}

@incollection{hart1988development,
  title={Development of NASA-TLX (Task Load Index): Results of empirical and theoretical research},
  author={Hart, Sandra G and Staveland, Lowell E},
  booktitle={Advances in psychology},
  volume={52},
  pages={139--183},
  year={1988},
  publisher={Elsevier}
}

@article{cohen2014focused,
  title={Focused and sustained attention},
  author={Cohen, Ronald A and Cohen, Ronald A},
  journal={The neuropsychology of attention},
  pages={89--112},
  year={2014},
  publisher={Springer}
}

@article{van2019heartpy,
  title={HeartPy: A novel heart rate algorithm for the analysis of noisy signals},
  author={Van Gent, Paul and Farah, Haneen and Van Nes, Nicole and Van Arem, Bart},
  journal={Transportation research part F: traffic psychology and behaviour},
  volume={66},
  pages={368--378},
  year={2019},
  publisher={Elsevier}
}

@ARTICLE{tremmel2019estimating,
  
AUTHOR={Tremmel, Christoph and Herff, Christian and Sato, Tetsuya and Rechowicz, Krzysztof and Yamani, Yusuke and Krusienski, Dean J.},   
	 
TITLE={Estimating Cognitive Workload in an Interactive Virtual Reality Environment Using EEG},      
	
JOURNAL={Frontiers in Human Neuroscience},      
	
VOLUME={13},           
	
YEAR={2019},      
	  
URL={https://www.frontiersin.org/articles/10.3389/fnhum.2019.00401},       
	
DOI={10.3389/fnhum.2019.00401},      
	
ISSN={1662-5161}
}

@article{hebbar2022cognitive,
  title={Cognitive load estimation in VR flight simulator},
  author={Hebbar, Archana and Vinod, Sanjana and Shah, Aumkar Kishore and Pashilkar, Abhay and Biswas, Pradipta},
  journal={Journal of Eye Movement Research},
  volume={15},
  number={3},
  year={2022}
}

@article{hanzal2023investigation,
  title={An investigation into discomfort and fatigue related to the wearing of an EEG neurofeedback headset},
  author={Hanzal, Simon and Tvrda, Lucie and Harvey, Monika},
  journal={medRxiv},
  pages={2023--02},
  year={2023},
  publisher={Cold Spring Harbor Laboratory Press}
}

@ARTICLE{brugada2022enhance,
  
AUTHOR={Brugada-Ramentol, Victòria  and Bozorgzadeh, Amir  and Jalali, Hossein },
         
TITLE={Enhance VR: A Multisensory Approach to Cognitive Training and Monitoring},
        
JOURNAL={Frontiers in Digital Health},
        
VOLUME={Volume 4 - 2022},

YEAR={2022},

URL={https://www.frontiersin.org/journals/digital-health/articles/10.3389/fdgth.2022.916052},

DOI={10.3389/fdgth.2022.916052},

ISSN={2673-253X}}

@article{willis2006long,
  title={Long-term effects of cognitive training on everyday functional outcomes in older adults},
  author={Willis, Sherry L and Tennstedt, Sharon L and Marsiske, Michael and Ball, Karlene and Elias, Jeffrey and Koepke, Karin M and Morris, John N and Rebok, George W and Unverzagt, Frederick W and Stoddard, Anne M and Wright, Elizabeth and {ACTIVE Study Group}},
  journal={JAMA},
  volume={296},
  number={23},
  pages={2805--2814},
  year={2006},
  publisher={American Medical Association},
  doi={10.1001/jama.296.23.2805}
}

@article{rebok2014ten,
  title={Ten-year effects of the advanced cognitive training for independent and vital elderly cognitive training trial on cognition and everyday functioning in older adults},
  author={Rebok, George W and Ball, Karlene and Guey, Lin T and Jones, Richard N and Kim, Hae-Young and King, Jonathan W and Marsiske, Michael and Morris, John N and Tennstedt, Sharon L and Unverzagt, Frederick W and Willis, Sherry L and {ACTIVE Study Group}},
  journal={Journal of the American Geriatrics Society},
  volume={62},
  number={1},
  pages={16--24},
  year={2014},
  publisher={Wiley},
  doi={10.1111/jgs.12607}
}

@article{yu2023effect,
  title={The effect of virtual reality on executive function in older adults with mild cognitive impairment: a systematic review and meta-analysis},
  author={Yu, Doris and Li, Xiuying and Lai, Florence Harmony},
  journal={Aging \& Mental Health},
  volume={27},
  number={4},
  pages={663--673},
  year={2023},
  publisher={Taylor \& Francis},
  doi={10.1080/13607863.2022.2076202}
}

@article{moulaei2024efficacy,
  title={Efficacy of virtual reality-based training programs and games on the improvement of cognitive disorders in patients: a systematic review and meta-analysis},
  author={Moulaei, Khadijeh and Sharifi, Hadi and Bahaadinbeigy, Kambiz and Bagherzadeh, Razieh and Haghdoost, Ali Akbar},
  journal={BMC Psychiatry},
  volume={24},
  number={1},
  pages={116},
  year={2024},
  publisher={Springer},
  doi={10.1186/s12888-024-05563-z}
}

@article{johansen2024virtual,
  title={Virtual reality as a method of cognitive training of processing speed, working memory, and sustained attention in persons with acquired brain injury: a protocol for a randomized controlled trial},
  author={Johansen, Torun and Matre, Marianne and Løvstad, Marianne and Lund, Anne and Martinsen, Anne Christin and Olsen, Alexander and Becker, Frank and Brunborg, Cathrine and Ponsford, Jennie and Spikman, Jacoba and Neumann, Dawn and Tornås, Solveig},
  journal={Trials},
  volume={25},
  number={1},
  pages={340},
  year={2024},
  publisher={Springer},
  doi={10.1186/s13063-024-08178-7}
}

@article{mcphee2022dual,
  title={Dual-task interference as a function of varying motor and cognitive demands},
  author={McPhee, Ashley M and Cheung, Tsz Ching Karen and Schmuckler, Mark A},
  journal={Frontiers in Psychology},
  volume={13},
  pages={952245},
  year={2022},
  publisher={Frontiers Media},
  doi={10.3389/fpsyg.2022.952245}
}

@article{ahmad2023framework,
  title={A framework to estimate cognitive load using physiological data},
  author={Ahmad, Muneeb Imtiaz and Keller, Ingo and Robb, David A and Lohan, Katrin S},
  journal={Personal and Ubiquitous Computing},
  volume={27},
  pages={2027--2041},
  year={2023},
  publisher={Springer},
  doi={10.1007/s00779-020-01455-7}
}

@article{muhl2014eeg,
  title={EEG-based workload estimation across affective contexts},
  author={M{\"u}hl, Christian and Jeunet, Camille and Lotte, Fabien},
  journal={Frontiers in Neuroscience},
  volume={8},
  pages={114},
  year={2014},
  publisher={Frontiers Media},
  doi={10.3389/fnins.2014.00114}
}

@article{anders2024unobtrusive,
  title={Unobtrusive measurement of cognitive load and physiological signals in uncontrolled environments},
  author={Anders, Christoph and Moontaha, Syeda and Real, Sebastian and Demberg, Vera and Lux, Mathias},
  journal={Scientific Data},
  volume={11},
  pages={1000},
  year={2024},
  publisher={Nature Publishing Group},
  doi={10.1038/s41597-024-03738-7}
}

@article{lagomarsino2022online,
  title={An Online Framework for Cognitive Load Assessment in Industrial Tasks},
  author={Lagomarsino, Marta and Lorenzini, Marta and De Momi, Elena and Ajoudani, Arash},
  journal={Robotics and Computer-Integrated Manufacturing},
  volume={78},
  pages={102380},
  year={2022},
  publisher={Elsevier},
  issn={0736-5845},
  doi={10.1016/j.rcim.2022.102380},
  url={https://doi.org/10.1016/j.rcim.2022.102380}
}

@article{kim2020hidden,
  title={A hidden Markov model for analyzing eye-tracking of moving objects},
  author={Kim, J. and Singh, S. and Thiessen, E. D. and Fisher, A. V.},
  journal={Behavior Research Methods},
  volume={52},
  pages={1225--1243},
  year={2020},
  publisher={Springer},
  doi={10.3758/s13428-019-01313-2},
  url={https://doi.org/10.3758/s13428-019-01313-2}
}

@article{asgher2020enhanced,
  title={Enhanced Accuracy for Multiclass Mental Workload Detection Using Long Short-Term Memory for Brain–Computer Interface},
  author={Asgher, U. and Khalil, K. and Khan, M. J. and Ahmad, R. and Butt, S. I. and Ayaz, Y. and Naseer, N. and Nazir, S.},
  journal={Frontiers in Neuroscience},
  volume={14},
  pages={584},
  year={2020},
  publisher={Frontiers Media},
  doi={10.3389/fnins.2020.00584}
}

@article{bachurina2022multiple,
title = {Multiple levels of mental attentional demand modulate peak saccade velocity and blink rate},
journal = {Heliyon},
volume = {8},
number = {1},
pages = {e08826},
year = {2022},
issn = {2405-8440},
doi = {https://doi.org/10.1016/j.heliyon.2022.e08826},
url = {https://www.sciencedirect.com/science/article/pii/S2405844022001141},
author = {Valentina Bachurina and Marie Arsalidou},
}

@article{behrens2021quantifying,
author = {Behrens, Friederike and Moulder, Robert and Boker, Steven and Kret, Mariska},
year = {2020},
month = {08},
pages = {},
title = {Quantifying Physiological Synchrony through Windowed Cross-Correlation Analysis: Statistical and Theoretical Considerations},
doi = {10.1101/2020.08.27.269746}
}

@article{wang2023characterisation,
  title={Characterisation of Cognitive Load Using Machine Learning Classifiers of Electroencephalogram Data},
  author={Wang, Q. and Smythe, D. and Cao, J. and Hu, Z. and Proctor, K. J. and Owens, A. P. and Zhao, Y.},
  journal={Sensors},
  volume={23},
  pages={8528},
  year={2023},
  publisher={MDPI},
  doi={10.3390/s23208528},
  url={https://doi.org/10.3390/s23208528}
}

@article{coutrot2018scanpath,
  title={Scanpath modeling and classification with hidden Markov models},
  author={Coutrot, A. and Hsiao, J. H. and Chan, A. B.},
  journal={Behavior Research Methods},
  volume={50},
  pages={362--379},
  year={2018},
  publisher={Springer},
  doi={10.3758/s13428-017-0876-8},
  url={https://doi.org/10.3758/s13428-017-0876-8}
}

@article{distasi2010saccadic,
  title={Saccadic peak velocity sensitivity to variations in mental workload},
  author={Di Stasi, L. L. and Renner, R. and Staehr, P. and Helmert, J. R. and Velichkovsky, B. M. and Cañas, J. J. and Catena, A. and Pannasch, S.},
  journal={Aviation, Space, and Environmental Medicine},
  volume={81},
  number={4},
  pages={413--417},
  year={2010},
  publisher={Aerospace Medical Association},
  doi={10.3357/asem.2579.2010},
  url={https://doi.org/10.3357/asem.2579.2010}
}

@article{sarailoo2022assessment,
  title={Assessment of instantaneous cognitive load imposed by educational multimedia using electroencephalography signals},
  author={Sarailoo, R. and Latifzadeh, K. and Amiri, S. H. and Bosaghzadeh, A. and Ebrahimpour, R.},
  journal={Frontiers in Neuroscience},
  volume={16},
  pages={744737},
  year={2022},
  publisher={Frontiers Media},
  doi={10.3389/fnins.2022.744737}
}

@inproceedings{kunjan2021necessity,
  title={The Necessity of Leave One Subject Out (LOSO) Cross Validation for EEG Disease Diagnosis},
  author={Kunjan, S. and others},
  booktitle={Brain Informatics},
  pages={558--567},
  year={2021},
  publisher={Springer},
  address={Cham},
  editor={Mahmud, M. and Kaiser, M. S. and Vassanelli, S. and Dai, Q. and Zhong, N.},
  series={Lecture Notes in Computer Science},
  volume={12960},
  doi={10.1007/978-3-030-86993-9_50},
  url={https://doi.org/10.1007/978-3-030-86993-9_50}
}

@article{lapitan2024estimation,
  title={Estimation of phase distortions of the photoplethysmographic signal in digital IIR filtering},
  author={Lapitan, D. G. and Rogatkin, D. A. and Molchanova, E. A. and others},
  journal={Scientific Reports},
  volume={14},
  pages={6546},
  year={2024},
  publisher={Nature Publishing Group},
  doi={10.1038/s41598-024-57297-3},
  url={https://doi.org/10.1038/s41598-024-57297-3}
}

@article{krejtz2018eye,
    doi = {10.1371/journal.pone.0203629},
    author = {Krejtz, Krzysztof AND Duchowski, Andrew T. AND Niedzielska, Anna AND Biele, Cezary AND Krejtz, Izabela},
    journal = {PLOS ONE},
    publisher = {Public Library of Science},
    title = {Eye tracking cognitive load using pupil diameter and microsaccades with fixed gaze},
    year = {2018},
    month = {09},
    volume = {13},
    url = {https://doi.org/10.1371/journal.pone.0203629},
    pages = {1-23},
    number = {9},
}

@article{larsen2024method,
  title={A method for synchronized use of EEG and eye tracking in fully immersive VR},
  author={Larsen, O. F. P. and Tresselt, W. G. and Lorenz, E. A. and Holt, T. and Sandstrak, G. and Hansen, T. I. and Su, X. and Holt, A.},
  journal={Frontiers in Human Neuroscience},
  volume={18},
  pages={1347974},
  year={2024},
  publisher={Frontiers Media},
  doi={10.3389/fnhum.2024.1347974}
}

@article{nystrom2024blink,
  title={What is a blink? Classifying and characterizing blinks in eye openness signals},
  author={Nyström, M. and Andersson, R. and Niehorster, D. C. and others},
  journal={Behavior Research Methods},
  volume={56},
  pages={3280--3299},
  year={2024},
  publisher={Springer},
  doi={10.3758/s13428-023-02333-9},
  url={https://doi.org/10.3758/s13428-023-02333-9}
}

@article{pham2021timefrequency,
  title={Time–frequency time–space LSTM for robust classification of physiological signals},
  author={Pham, T. D.},
  journal={Scientific Reports},
  volume={11},
  pages={6936},
  year={2021},
  publisher={Nature Publishing Group},
  doi={10.1038/s41598-021-86432-7},
  url={https://doi.org/10.1038/s41598-021-86432-7}
}

@article{raju2021filtering,
  title={Filtering Eye-Tracking Data From an EyeLink 1000: Comparing Heuristic, Savitzky-Golay, IIR and FIR Digital Filters},
  author={Raju, M. H. and Friedman, L. and Bouman, T. M. and Komogortsev, O. V.},
  journal={Journal of Eye Movement Research},
  volume={14},
  number={3},
  pages={1--16},
  year={2021},
  doi={10.16910/jemr.14.3.6},
  url={https://doi.org/10.16910/jemr.14.3.6}
}

@article{solhjoo2019heart,
  title={Heart Rate and Heart Rate Variability Correlate with Clinical Reasoning Performance and Self-Reported Measures of Cognitive Load},
  author={Solhjoo, S. and Haigney, M. C. and McBee, E. and others},
  journal={Scientific Reports},
  volume={9},
  pages={14668},
  year={2019},
  publisher={Nature Publishing Group},
  doi={10.1038/s41598-019-50280-3},
  url={https://doi.org/10.1038/s41598-019-50280-3}
}

@article{zhang2022mental,
  title={A Mental Workload Evaluation Model Based on Improved Multibranch LSTM Network with Attention Mechanism},
  author={Zhang, Hongning},
  journal={Advances in Multimedia},
  volume={2022},
  pages={9601946},
  year={2022},
  publisher={Hindawi},
  doi={10.1155/2022/9601946},
  url={https://doi.org/10.1155/2022/9601946}
}

@article{zanetti2022real,
  title={Real-Time EEG-Based Cognitive Workload Monitoring on Wearable Devices},
  author={Zanetti, R. and Arza, A. and Aminifar, A. and Atienza, D.},
  journal={IEEE Transactions on Biomedical Engineering},
  volume={69},
  number={1},
  pages={265--277},
  year={2022},
  publisher={IEEE},
  doi={10.1109/TBME.2021.3092206},
  url={https://doi.org/10.1109/TBME.2021.3092206}
}

@article{zangroniz2017observing,
    doi = {10.1371/journal.pone.0003082},
    author = {Hägni, Karin AND Eng, Kynan AND Hepp-Reymond, Marie-Claude AND Holper, Lisa AND Keisker, Birgit AND Siekierka, Ewa AND Kiper, Daniel C.},
    journal = {PLOS ONE},
    publisher = {Public Library of Science},
    title = {Observing Virtual Arms that You Imagine Are Yours Increases the Galvanic Skin Response to an Unexpected Threat},
    year = {2008},
    month = {08},
    volume = {3},
    url = {https://doi.org/10.1371/journal.pone.0003082},
    pages = {1-6},
    number = {8},

}

@INPROCEEDINGS{omnicept,
  author={Wei, Jishang and Siegel, Erika and Sundaramoorthy, Prahalathan and Gomes, Antônio and Zhang, Shibo and Vankipuram, Mithra and Smathers, Kevin and Ghosh, Sarthak and Horii, Hiroshi and Bailenson, Jeremy and Ballagas, Rafael 'Tico'},
  booktitle={2025 IEEE Conference Virtual Reality and 3D User Interfaces (VR)}, 
  title={Cognitive Load Inference Using Physiological Markers in Virtual Reality}, 
  year={2025},
  volume={},
  number={},
  pages={759-769},
  doi={10.1109/VR59515.2025.00098}
}

@article{clare2025,
author = {Bhatti, Anubhav and Angkan, Prithila and Behinaein, Behnam and Mahmud, Zunayed and Rodenburg, Dirk and Braund, Heather and Mclellan, P. and Ruberto, Aaron and Harrison, Geoffery and Wilson, Daryl and Szulewski, Adam and Howes, Dan and Etemad, Ali and Hungler, P.},
year = {2025},
month = {01},
pages = {1-13},
title = {CLARE: Cognitive Load Assessment in Real-time with Multimodal Data},
volume = {PP},
journal = {IEEE Transactions on Cognitive and Developmental Systems},
doi = {10.1109/TCDS.2025.3555517}
}

@InProceedings{araujo2019,
author="Araujo, Vladimir
and Gonzalez, Alejandra
and Mendez, Diego",
editor="Botto-Tobar, Miguel
and Pizarro, Guillermo
and Z{\'u}{\~{n}}iga-Prieto, Miguel
and D'Armas, Mayra
and Z{\'u}{\~{n}}iga S{\'a}nchez, Miguel",
title="Dynamic Difficulty Adjustment for a Memory Game",
booktitle="Technology Trends",
year="2019",
publisher="Springer International Publishing",
address="Cham",
pages="605--616",
isbn="978-3-030-05532-5"
}

@inproceedings{drey2020,
author = {Drey, Tobias and Jansen, Pascal and Fischbach, Fabian and Frommel, Julian and Rukzio, Enrico},
title = {Towards Progress Assessment for Adaptive Hints in Educational Virtual Reality Games},
year = {2020},
isbn = {9781450368193},
publisher = {Association for Computing Machinery},
address = {New York, NY, USA},
url = {https://doi.org/10.1145/3334480.3382789},
doi = {10.1145/3334480.3382789},
booktitle = {Extended Abstracts of the 2020 CHI Conference on Human Factors in Computing Systems},
pages = {1–9},
numpages = {9},
location = {Honolulu, HI, USA},
series = {CHI EA '20}
}

@article{Salminen2024,
author = {Mikko Salminen and Simo Järvelä and Ilkka Kosunen and Antti Ruonala and Juho Hamari and Niklas Ravaja and Giulio Jacucci},
title = {Meditating in a neurofeedback virtual reality: effects on sense of presence, meditation depth and brain oscillations},
journal = {Behaviour \& Information Technology},
volume = {43},
number = {12},
pages = {2750--2764},
year = {2024},
publisher = {Taylor \& Francis},
doi = {10.1080/0144929X.2023.2258231},


URL = { 
    
        https://doi.org/10.1080/0144929X.2023.2258231
    
    

},
eprint = { 
    
        https://doi.org/10.1080/0144929X.2023.2258231
    
    

}
}

@INPROCEEDINGS{szczepaniak2024mldriven,
  author={Szczepaniak, Dominik and Harvey, Monika and Deligianni, Fani},
  booktitle={2024 IEEE International Symposium on Mixed and Augmented Reality Adjunct (ISMAR-Adjunct)}, 
  title={ML-Driven Cognitive Workload Estimation in a VR-based Sustained Attention Task*}, 
  year={2024},
  volume={},
  number={},
  pages={557-560},
  doi={10.1109/ISMAR-Adjunct64951.2024.00160}}

@article{chiossi2023adapting,
  title={Adapting Visual Complexity Based on Electrodermal Activity Improves Working Memory Performance in Virtual Reality},
  author={Chiossi, Francesco and Turgut, Yaz and Welsch, Robin and Mayer, Sven},
  journal={Proceedings of the ACM on Human-Computer Interaction},
  volume={7},
  number={MHCI},
  pages={1--26},
  year={2023},
  publisher={ACM}
}

@inproceedings{lindlbauer2019context,
  title={Context-Aware Online Adaptation of Mixed Reality Interfaces},
  author={Lindlbauer, David and Feit, Anna Maria and Hilliges, Otmar},
  booktitle={Proceedings of the 32nd Annual ACM Symposium on User Interface Software and Technology},
  pages={147--160},
  year={2019},
  publisher={ACM}
}

@article{pinilla2023real,
  title={Real-time affect detection in virtual reality: a technique based on a three-dimensional model of affect and {EEG} signals},
  author={Pinilla, A. and Voigt-Antons, J.-N. and Garcia, J. and Raffe, W. and M{\"o}ller, S.},
  journal={Frontiers in Virtual Reality},
  volume={3},
  pages={1088647},
  year={2023},
  publisher={Frontiers}
}

@article{aranha2021adapting,
  title={Adapting Software with Affective Computing: A Systematic Review},
  author={Aranha, R. V. and Correa, C. G. and Nunes, F. L. S.},
  journal={IEEE Transactions on Affective Computing},
  volume={12},
  number={4},
  pages={883--899},
  year={2021},
  publisher={IEEE}
}

@article{zahabi2020adaptive,
  title={Adaptive virtual reality-based training: a systematic literature review and framework},
  author={Zahabi, M. and Abdul Razak, A. M.},
  journal={Virtual Reality},
  volume={24},
  number={4},
  pages={725--752},
  year={2020},
  publisher={Springer}
}

@article{quintero2025personalized,
  title={Personalized Feature Importance Ranking for Affect Recognition From Behavioral and Physiological Data},
  author={Quintero, L. and Fors, U. and Papapetrou, P.},
  journal={IEEE Transactions on Games},
  volume={17},
  number={3},
  pages={594--603},
  year={2025},
  publisher={IEEE}
}

@article{motes2020multidomain,
  title={Multidomain Cognitive Training Transfers to Attentional and Executive Functions in Healthy Older Adults},
  author={Motes, Michael A. and Gamino, Jacquelyn F. and Chapman, Sandra B. and Rao, Nadh K. and Maguire, Mandy J. and Brier, Matthew R. and Kraut, Michael A. and Hart Jr, John},
  journal={Frontiers in Human Neuroscience},
  volume={14},
  pages={586963},
  year={2020},
  publisher={Frontiers},
  doi={10.3389/fnhum.2020.586963}
}

@article{cassidy2024training,
  title={Training of Executive Functions in Children: A Meta-Analysis of Cognitive Training Interventions},
  author={Cassidy, Sarah J. and Roche, Bryan and Colbert, Dylan},
  journal={SAGE Open},
  volume={14},
  number={2},
  year={2024},
  publisher={SAGE Publications},
  doi={10.1177/21582440241311060}
}

@article{mortazavi2024dynamic,
  title={Dynamic difficulty adjustment approaches in video games: a systematic literature review},
  author={Mortazavi, Fatemeh and Moradi, Hadi and Vahabie, Abdol-Hossein},
  journal={Multimedia Tools and Applications},
  volume={83},
  pages={83227--83274},
  year={2024},
  publisher={Springer},
  doi={10.1007/s11042-024-18768-x}
}

@article{yeh1988dissociation,
  title={Dissociation of performance and subjective measures of workload},
  author={Yeh, Yei-Yu and Wickens, Christopher D},
  journal={Human Factors},
  volume={30},
  number={1},
  pages={111--120},
  year={1988},
  publisher={SAGE Publications},
  doi={10.1177/001872088803000110}
}

@article{hendy1993measuring,
  title={Measuring subjective workload: When is one scale better than many?},
  author={Hendy, Keith C and Hamilton, Kevin M and Landry, Lois N},
  journal={Human Factors},
  volume={35},
  number={4},
  pages={579--601},
  year={1993},
  publisher={SAGE Publications},
  doi={10.1177/001872089303500401}
}

@article{guo2025exploratory,
  title={An exploratory study of the relationship between objective game difficulty and subjective game difficulty},
  author={Guo, Zhixing and Ren, Xiangshi},
  journal={International Journal of Human-Computer Studies},
  volume={199},
  pages={103502},
  year={2025},
  publisher={Elsevier},
  doi={10.1016/j.ijhcs.2024.103502}
}

@inproceedings{constant2019dynamic,
  title={Dynamic Difficulty Adjustment Impact on Players' Confidence},
  author={Constant, Thomas and Levieux, Guillaume},
  booktitle={Proceedings of the 2019 CHI Conference on Human Factors in Computing Systems},
  pages={1--12},
  year={2019},
  publisher={ACM},
  address={Glasgow, Scotland UK},
  doi={10.1145/3290605.3300693}
}

@article{guo2024rethinking,
  title={Rethinking dynamic difficulty adjustment for video game design},
  author={Guo, Zhixing and Thawonmas, Ruck and Ren, Xiangshi},
  journal={Entertainment Computing},
  volume={50},
  pages={100663},
  year={2024},
  publisher={Elsevier},
  doi={10.1016/j.entcom.2024.100663}
}

@inproceedings{constant2017objective,
  title={From Objective to Subjective Difficulty Evaluation in Video Games},
  author={Constant, Thomas and Levieux, Guillaume and Buendia, Axel and Natkin, St{\'e}phane},
  booktitle={Entertainment Computing -- ICEC 2017},
  series={Lecture Notes in Computer Science},
  volume={10507},
  pages={3--10},
  year={2017},
  publisher={Springer},
  doi={10.1007/978-3-319-67684-5_1}
}



\end{document}